\documentclass[useAMS,usenatbib]{mn2e}
\usepackage{graphicx}
\usepackage{longtable}
\usepackage{natbib}
\def\aj{AJ}%
\def\apj{ApJ}%
\def\apjl{ApJ}%
\def\apjs{ApJS}%
\def\aap{A\&A}%
\def\aaps{A\&AS}%
\def\mnras{MNRAS}%
\def\nat{Nature}%
\title[Probing the submillimetre number counts at $f_{850{\rm \mu m}} < 2$\,mJy]{Probing the submillimetre number counts at $f_{850{\rm \mu m}} < 2$\,mJy}
\author[K.K.~Knudsen, P.P.~van~der~Werf, J.-P.~Kneib]{K.K.~Knudsen$^{1}$\thanks{E-mail: knudsen@astro.uni-bonn.de}, P.P.~van~der~Werf$^{2}$, J.-P.~Kneib$^{3,4,5}$\\
$^{1}$ Argelander Institute for Astronomy, University of Bonn, Auf dem H\"ugel 71, D-53121 Bonn, Germany \\
$^{2}$ Leiden Observatory, Leiden University, P.O.~Box 9513, NL--2300 RA Leiden, The Netherlands  \\
$^{3}$ Observatoire Midi-Pyr\'en\'ees, UMR5572,
     14 Avenue Edouard Belin, 31000 Toulouse, France\\
$^{4}$ Caltech, Astronomy, 105-24, Pasadena, CA 91125, USA \\
$^{5}$ OAMP, Laboratoire d'Astrophysique de Marseille, traverse du Siphon, 13012 Marseille, France}

\begin{document}

\date{Accepted 2007 December 5.  Received 2007 October 24; in original form 2007 January 14}

\pagerange{\pageref{firstpage}--\pageref{lastpage}} \pubyear{2007}

\maketitle

\label{firstpage}

%%%%%%%%%%%%%%%%%%%%%%%%%%%%%%%%%%%%%%%%%%%%%%%%%%%%%%%%%%%%%%%%%%
%%%    ABSTRACT
%%%%%%%%%%%%%%%%%%%%%%%%%%%%%%%%%%%%%%%%%%%%%%%%%%%%%%%%%%%%%%%%%%
\begin{abstract}
We have conducted a submillimetre mapping survey of faint, gravitationally 
lensed sources, where we have targeted twelve galaxy clusters and additionally 
the NTT Deep Field. 
The total area surveyed is 71.5\,arcmin$^2$ in the image plane;  correcting 
for gravitational lensing, the total area surveyed is 40\,arcmin$^2$ in the 
source plane for a typical source redshift $z\approx $ 2.5. 
In the deepest maps, an image plane depth of 1$\sigma$ r.m.s.\ $\sim $ 
0.8\,mJy is reached. 
This survey is the largest survey to date to reach such depths.  
In total 59 sources were detected, including three multiply-imaged 
sources.  
The gravitational lensing makes it possible to detect sources with flux 
density below the blank field confusion limit. 
The lensing corrected fluxes ranges from 0.11 mJy to 19 mJy.  
After correcting for multiplicity there are 10 sources with fluxes 
$< 2$\,mJy of which 7 have sub-mJy fluxes, 
doubling the number of such sources known. 
Number counts are determined below the confusion limit. 
At 1\,mJy the integrated number count is $\sim $ 10$^{4}$\,deg$^{-2}$, 
and at 0.5\,mJy it is $\sim $ 2$\times$10$^{4}$\,deg$^{-2}$. 
Based on the number counts, at a source plan flux limit of 0.1 mJy, 
essentially all of the 850\,$\mu$m background emission has been resolved. 
The dominant contribution ($> 50$ per cent) to the integrated background
arises from sources with fluxes $S_{\it 850}$ between 0.4 and 2.5\,mJy, 
while the bright sources $S_{850} > $ 6\,mJy contribute only 10 per cent. 
\end{abstract}

\begin{keywords}
Survey -- submillimetre -- galaxies: starburst -- galaxies: high redshift -- galaxies: evolution
\end{keywords} 

% ====================================================================

%%%%%%%%%%%%%%%%%%%%%%%%%%%%%%%%%%%%%%%%%%%%%%%%%%%%%%%%%%%%%%%%%%
%%%    INTRODUCTION
%%%%%%%%%%%%%%%%%%%%%%%%%%%%%%%%%%%%%%%%%%%%%%%%%%%%%%%%%%%%%%%%%%

\section{Introduction}
\label{sec:intro}

The first submillimetre mapping instrument SCUBA \citep[Submillimetre Common 
User Bolometer Array;][]{holland99}, which is mounted at 
the James Clerk Maxwell Telescope (JCMT) at Hawaii, allowed for observations of 
IR luminous galaxies at high redshift.  The first observations at 
850\,$\mu$m \citep{smail97} 
showed that these objects are much more common at earlier 
epochs.  Subsequently, a number of surveys have been undertaken to study this 
population of submm detected galaxies.  The blank field surveys 
include observations of the Hubble Deep Field North (HDF-N) 
\citep{hughes98,borys03,serjeant03}, the Hawaii Deep Fields \citep{barger99}, 
Canada-UK Deep SCUBA Survey (CUDSS) \citep{eales00,webb_3h}, 
the 8\,mJy survey \citep{scott_8mJy}, Galactic regions \citep{barnard04}, 
the Groth strip \citep{coppin05}, the SHADES survey \citep{coppin06}, 
and a re-analysis of several blank field surveys \citep{scott06}.
In particular CUDSS and SHADES have been successful in 
covering a large area of the sky.  However, the blank field surveys are 
limited by the confusion at 2\,mJy at 850\,$\mu$m with the 15\,m JCMT 
and hence do not probe the number counts of the fainter population.  
The submillimetre extragalactic background light \citep{fixsen98,puget96} is, 
however, dominated by the population around 1\,mJy \citep[e.g.][]{barger99,blain99b,cowie02}. 
To break the blank field confusion limit observing with SCUBA, gravitational 
lensing must be employed.  The UK-SCUBA Lens Survey \citep{smail97,sibk02}
targeted seven galaxy cluster fields.  Three of their fields were 
observed to larger depth \citep{cowie02}.  
Another lens survey was performed by \citet{chapman02}, 
however, this survey was relatively shallow.   

Submillimetre observations of objects at high redshifts, $z> 1$, 
benefit from the fact that the geometrical dimming
of the light is cancelled by the negative $k$-correction, 
resulting from the fact that the peak of the spectral energy distribution
(SED) is shifted toward the observing band.   
For a given luminosity, the observed submm flux is close to constant 
between redshift 1 and 8. 
Consequently, extragalactic submm observations primarily probe the 
high redshift universe.  
Furthermore, deeper surveys do 
not probe deeper into the universe, but only sample lower luminosity 
galaxies. 
Galaxies in clusters at redshifts $z < 1$ are not expected to be seen with 
SCUBA, except for sometimes the central cD galaxy \citep{edge99}
or an Active Galactic Nucleus (AGN). 

We here present the Leiden-SCUBA Lens Survey, in which we have targeted 
twelve galaxy clusters.  
This is the largest survey so far of gravitationally lensing clusters, and 
it is the first survey to substantially probe below the blank field 
confusion limit. 
This paper presents the observations, the analysis 
of the data, the resulting catalogue and the number counts.  
The analysis involves the mathematically rigorous Mexican Hat 
wavelets algorithm \citep[e.g.][]{cayon00} and Monte Carlo simulations.  
In following papers, we will use the derived number counts as an 
observational constraint on models of the submm galaxy population, and 
we will present multiwavelength follow-up observations.

In Section~\ref{sec:obs_red} we present the observations and the 
reduction of the data.  
The source extraction is discussed in detail in Section \ref{sec:src_ext}.  
The issue of confusion is discussed in Section \ref{sec:confusion}, 
and the effect of gravitational lensing is discussed in Section 
\ref{sec:lensing}.   The resulting catalogue 
is presented in Section \ref{sec:catalogue}.  
Finally, in Section \ref{sec:counts} we present the number counts 
for the survey.  
Throughout the paper we assume $\Omega_m = 0.3$, $\Omega_{\Lambda} = 0.7$ 
and $H_0 = 70\,{\rm km/s/Mpc}$.

%%%%%%%%%%%%%%%%%%%%%%%%%%%%%%%%%%%%%%%%%%%%%%%%%%%%%%%%%%%%%%%%%%
%%%    OBSERVATIONS AND REDUCTION
%%%%%%%%%%%%%%%%%%%%%%%%%%%%%%%%%%%%%%%%%%%%%%%%%%%%%%%%%%%%%%%%%%

\section{Observations and reduction}
\label{sec:obs_red}

We have obtained observations of a number of clusters of galaxies 
at 850\,$\mu$m and 450\,$\mu$m with SCUBA. 
In addition we have obtained similar observations of the 
NTT Deep Field \citep{arnouts}, which was chosen 
due to the large, deep data set existing at optical and near-infrared 
wavelengths. 
In total our survey contains twelve fields of galaxy clusters
and the one blank field covering an area of 71.5\,arcmin$^2$. 
The parameters for each field are listed in Table~\ref{tab:obs_red}. 
\begin{table*}%[]
\caption[]{The observed fields. 
\label{tab:obs_red}}
\begin{center}
\begin{tabular}{lc@{\hspace{2mm}}c@{\hspace{2mm}}cc@{\hspace{2mm}}c@{\hspace{2mm}}cll@{\hspace{-1mm}}c@{\hspace{2mm}}c@{\hspace{2mm}}c@{\hspace{2mm}}c@{\hspace{2mm}}c}
\hline
\hline \\ [-2mm]
{\sc Name} & \multicolumn{3}{c}{RA(J2000)} & \multicolumn{3}{c}{Dec(J2000)}  &  \multicolumn{1}{c}{$z_{cl}$} & $t_{int}$ & $\Omega_{850}$ & $\sigma_{deep}^{850}$ & $\sigma_{wghtd}^{850}$ & $\sigma_{deep}^{450}$ & $\sigma_{wghtd}^{450}$ \\ 
& h&m&s & $^\circ$&$'$&$''$ & & hours & arcmin$^2$ & mJy/$\Omega_b$ & mJy/$\Omega_b$ & mJy/$\Omega_b$ & mJy/$\Omega_b$ \\ [2mm]
\hline \\ [-3mm] 
Cl0016+16$^{(\star)}$      & 00&18&33.2 & $+$16&26&17.8   & 0.541   &  7.73(5.46)    & 4.5  & 1.33 & 2.00   &  9.8 & 16.5 \\
A478$^{(\star)}$           & 04&13&25.3 & $+$10&27&54.3   & 0.0881  &  7.08          & 4.3  & 1.59 & 2.05   &  9.1 & 14.5 \\
A496           & 04&33&37.8 & $-$13&15&43.0  & 0.0328 & 10.4           & 4.1  & 1.08 & 1.47   & 11.6 & 17.2 \\
A520           & 04&54&07.0 & $+$02&55&12.0   & 0.202   &  19.6(18.0)    & 4.3  & 0.97 & 1.26   &  9.2 & 14.5 \\
MS1054$-$03 & 10&56&56.1 & $-$03&36&26.0  & 0.826   &  49.2         & 14.4 & 0.86 & 1.49   &  3.7 & 10.2 \\
A1689$^{(\star)}$          & 13&11&17.0 & $-$01&20&29.0  & 0.181   &  33.4(32.2)     & 5.4  & 0.70 & 0.97   &  4.4 &  9.9 \\
RXJ1347.5$-$1145$^{(\star)}$ & 13&47&30.5 & $-$11&45&09.0  & 0.451   &  10.5         & 4.8  & 2.04 & 3.06   &  7.2 & 24.8 \\
MS1358+62      & 13&59&50.6 & $+$62&31&05.1   & 0.328   &  4.80          & 4.2  & 1.39 & 1.81   &  7.6 & 11.2 \\
A2204$^{\star}$          & 16&32&46.9 & $+$05&34&33.0   & 0.1523  &  1.60          & 4.0  & 3.75 & 5.20   & 65.3 & 87.2 \\
A2218          & 16&35&54.2 & $+$66&12&37.0   & 0.171   &  42.3(35.6)    & 7.7  & 0.65 & 1.06   &  3.2 & 16.0 \\
A2219$^{(\star)}$          & 16&40&20.4 & $+$46&42&59.0   & 0.225   &  9.63          & 4.6  & 1.10 & 1.54   &  6.7 & 11.8 \\
A2597          & 23&25&19.8 & $-$12&07&26.4  & 0.0852  &  6.76           & 4.1  & 1.34 & 1.77   & 11.2 & 17.8 \\  [1mm]
NTT Deep Field & 12&05&22.6 & $-$07&44&14.9  & --na--  &  27.1           & 5.0  & 0.78 & 0.97   &  3.9 &  6.3 \\ [1mm]
\hline
\end{tabular}
\flushleft{Parameters of the observed fields.  
The integration time $t_{int}$ is the total integration time, but without 
overheads (i.e., without the time needed for jiggle, chopping, etc). 
If the 450\,$\mu$m exposure time is different from the 850\,$\mu$m
its value is given in parentheses. 
The area, $\Omega_{850}$, given is the field covered after removing the noisy edge.  
$\sigma_{deep}$ is the lowest noise value in the whole field.  
$\sigma_{wghtd}$ is the area-weighted noise level of the field. 
$\Omega_b$ is the beam. 
\\ 
$^{\star}$ Data from the JCMT archive.  $^{(\star)}$ Supplemented with data from the JCMT archive.
}
\end{center}
\end{table*}

SCUBA has two arrays of 37 and 91 bolometers optimized for 850\,$\mu$m 
respectively 450\,$\mu$m.  A dichroic beamsplitter is used for 
simultaneous observations with both arrays. 
Both arrays have the bolometers arranged in a hexagonal pattern.  Because 
SCUBA does not have a field rotator, the arrays appear as rotating 
on the sky. 
The field--of--view on the sky, which is approximately the same for 
both arrays, is roughly circular with a diameter of 2.3'. 
The observations were carried out in jiggle mode with a 64 point jiggle 
pattern, in order to fully sample the beam at both operating wavelengths. 
Subtraction of the strong sky background was done through 7.8\,Hz chopping 
with the secondary mirror. 
Our observations were performed with a chop throw of 45$''$ with the chopping 
position angle fixed in right ascension (RA). As a result the beam pattern 
has a central positive peak 
with negative sidelobes on each side, each with minus half the peak value, 
a pattern which can be used for the detection of at least the brighter sources. 
During the observations the pointing was checked every hour by observing 
bright blazars near the targeted fields.  The noise level of the arrays was 
checked at least twice during an observing shift, and the atmospheric opacity, 
$\tau$, was determined with JCMT at 850\,$\mu$m and 450\,$\mu$m 
every two--three hours and supplemented with the $\tau_{225{\rm GHz}}$ data 
from the neighboring Caltech Submillimetre Observatory (CSO). 
Calibrators were observed every two to three hours.  If available, primary 
calibrators, i.e.\ planets, preferably Uranus, were observed at least once 
during an observing shift. 
Our observations were supplemented with archival SCUBA data, hence the 
data set includes twelve cluster fields and the NTT Deep Field. 

The data were reduced using the {\sc surf} package \citep{surf}. 
First the chop of the secondary mirror was removed, i.e.\ the off--source 
measurements were subtracted from the on--source measurements.  
Then the varying responses of 
the bolometers were corrected by dividing with the array's flatfield. 
The extinction correction was performed based on the atmospheric opacity 
measured both with the JCMT and CSO. 
The $\tau_{850\mu{\rm m}}$ and $\tau_{450\mu{\rm m}}$ was measured a number 
of times during the night with the JCMT.  
As the atmospheric opacity may change on shorter timescales, the 
interpolated $\tau$-values may be somewhat inaccurate.  
At the CSO on the other hand, 
the opacity is measured every few minutes at 225 GHz.  
Using the linear relations between $\tau_{225{\rm GHz}}$ and $\tau_{850\mu
{\rm m}}$ respectively $\tau_{450\mu{\rm m}}$, deduced by \citet{archibald00},
it is possible to determine the atmospheric opacity at 
the time of the observations.  The zenith opacity was for most of the time 
\mbox{0.12 $ < \tau_{850\mu{\rm m}} < $ 0.40}.  
The data were inspected for bad or useless data.  For each scan for each 
bolometer datapoints deviating by more than three sigma, based on the 
r.m.s. of the individual bolometer,  were rejected from further analysis.  
This statistical exclusion is 
possible because there are no bright sources present in the data. 
Furthermore,  the data were inspected by eye, and bolometers that were 
clearly more noisy than other bolometers were flagged and excluded from 
further analysis. 
The pointing of each scan was corrected using the pointing observations 
taken just before and after the observations.  The r.m.s.\ pointing error 
of the JCMT is typically 2$''$.
The correlated atmospheric fluctuations still present in the data were 
subtracted using the pixel--by--pixel median of the 25 least noisy bolometers.

Data taken after January 2002, were affected by a periodic noise of 
currently unknown origin.  This was first pointed out by \citep{borys04}, 
and has later been discussed by \citet{webb05}, \citet{sawicki05}, 
and \citet{coppin06}. 
For these data, the 
power spectrum of the individual bolometers showed that some bolometers 
had a spike around 1/16s. 
This spike, however, did not systematically occur in the same bolometers or 
with the same strength.  
This effect was corrected for by performing a sky subtraction based on 
the bolometers that were not affected by this, and additionally, the 
affected bolometers were corrected through 
multiple-linear regression (T.\ Webb, private communication).  
As only a small fraction of the data for the survey was obtained after 
January 2002, this is relevant mostly for the MS1054-03 data, where the 
data most affected was the Northern pointing.  
The correction for the 1/16s spike 
brought down the noise in the affected data by $\sim$\,10 per cent. 

The scans were calibrated by multiplying by the 
flux conversion factors (FCF) determined from the calibration maps.  The 
FCFs are determined from the peak values (or the values corrected for 
extendedness) of the used calibrators.  
We estimate the uncertainty in the flux 
calibration as approximately 10 per cent at 850\,$\mu$m. 
At 450\,$\mu$m, the calibration uncertainty is about 30 per cent, 
because of variations in the beam profile resulting from thermal 
deformations of the dish.  This is in agreement with the canonical 
calibration uncertainties. 
The data were despiked by projecting the data on a grid and at each map pixel 
rejecting the associated bolometer pixels deviating by more than three sigma. 
Finally, all the bolometers were weighted based on their measured r.m.s. 
noise relative to one another and to the whole data set.  
The data were regridded with 1$''$ pixels into a final map.  
The beam sizes are $\sim$\,14.3$''$ at 850\,$\mu$m respectively 
$\sim$\,7.5$''$ at 450\,$\mu$m.  
The noisy edge was trimmed by removing the outer 23 pixels; 23 pixels 
corresponds to one and a half beam. 
Unless otherwise mentioned the maps used in the analysis 
have been smoothed with a 5$''$ full width half maximum (FWHM) Gaussian 
to reduce high spatial frequency noise, 
resulting in beam sizes of 15.1$''$ at 850\,$\mu$m and 9$''$ 
at 450\,$\mu$m.

%%%%%%%%%%%%%%%%%%%%%%%%%%%%%%%%%%%%%%%%%%%%%%%%%%%%%%%%%%%%%%%%%%
%%%    SOURCE EXTRACTION
%%%%%%%%%%%%%%%%%%%%%%%%%%%%%%%%%%%%%%%%%%%%%%%%%%%%%%%%%%%%%%%%%%

\section{Source extraction}
\label{sec:src_ext}

The sensitivity in the reduced maps is not uniform across the field.  
As we are working close to the signal--to--noise limit of these data 
it is crucial to understand the properties of the noise.  Furthermore, 
the source extraction from maps like the SCUBA maps is a non--trivial 
task, which must be performed with a robust and well--understood method.  
In a previous paper \citep{knudsen06}, we describe the approach, which 
we adopt for this survey and first applied to the data for the field 
A2218.  
In summary, the noise is measured across a map using 
Monte Carlo simulations, where the real data (i.e.\ the time streams) 
are substituted by the output from a random number generator with a 
Gaussian distribution and the same statistical properties as the 
real data. 
The simulated data is reduced following the same procedure as the real 
data, creating empty maps, and the standard deviation is measured 
for each pixel using about 500 simulated maps.   
Furthermore, in order to remove the chopping pattern (caused by the 
motion of the secondary mirror) from the beam we use the CLEAN 
algorithm \citep{hogbom74}.  For details, see \citet{knudsen06}.

\subsection{Point source detection}
\label{subsec:wavelets}

Most sources at high redshifts have an angular extent on the sky of less 
than a few arcseconds.  In the 850\,$\mu$m beam such sources will
appear as point sources.  To do the point source extraction we choose 
to use Mexican Hat Wavelets (MHW). MHW is a mathematically rigorous 
tool for which the performance can be fully understood and quantified. 
MHW has proven to be a powerful source extraction technique in both SCUBA
jiggle-maps and scan-maps \citep{barnard04,knudsen06}. 
Wavelets are mathematical functions used for analysing according to scale.  
Isotropic wavelets have the 
advantage that no assumptions about the underlying field need to be made. 
The beam at 850\,$\mu$m is well-described by a 2-dimensional 
Gaussian, which is best detected with the 'Mexican Hat' wavelets; 
the 'Mexican Hat' is the second derivative of a Gaussian.  
The software utilized are programmes written for the anticipated 
{\it Planck Surveyor}\footnote{http://www.esa.int/science/planck} 
mission, but modified for application on SCUBA maps.  
For more details and the application of the programmes on 
SCUBA maps see \citet{cayon00,barnard04,knudsen06}. 
Here we summarize only the relevant details. 

The MHW source extraction is done in the following way.  
Point source candidates are selected at positions with wavelet 
coefficient values larger than a given number.  For each candidate, 
the wavelet function is compared to the theoretical variations 
expected with the scale, as a further check on the source's shape. 
A $\chi^2$ value is calculated between the expected and the 
experimental results.  If the $\chi^2$ is smaller than a given 
limit, i.e., the region surrounding the identified peak 
has the characteristics of a Gaussian point source of the correct 
dimensions, the point source is included in the extracted catalogue.  
In \citet{knudsen06} we have performed controlled 
detection experiments to determine the optimal criteria for the 
MHW algorithm applied to SCUBA jiggle maps: the wavelet coefficient 
$\geq 2$ and $\chi^2 \leq 4$.    
These we combine with the $S/N \geq 3$ criterion in the flux map. 

In doing the source extraction, the MHW algorithm searches the maps for 
features at the scale of the beam, features with scales smaller or 
larger than the beam are not extracted.  E.g., in the field 
RX\,J1347.5$-$1145 the extended emission presumably due to the 
Sunyaev-Zel'dovich effect \citep{komatsu99} is undetected by MHW point 
source detection. 
Also extended sources cannot be detected by changing the $\chi^2$-limit. 
This of course ensures that all sources detected are point sources. 

To check for undetected sources located closer than one beam to a 
detected source, the detected sources are subtracted from the map.  
If those are the only point sources present 
in the map, then MHW finds no sources in the residual map.  If any 
sources are significantly detected in the residual map, those sources 
are subtracted from the original map and a new MHW detection is done 
in order to improve the result on the first detection.  This is 
continued iteratively until the output converges. 
This approach does not resolve extended sources. 

As described in \citet{knudsen06}, we performed detection experiments doing 
Monte Carlo simulations to analyse and determine the accuracy of the 
derived parameters (position and flux). 
In the simulations point sources are added on to empty maps and 
recovered using the MHW algorithm.  We perform this for all fields
and find similar results.  The accuracies derived through this are 
included in the final errors quoted in Table \ref{tab:catalogue}.  

\subsection{Completeness}
\label{subsec:complt}

\begin{figure}
\center{\includegraphics[width=7cm]{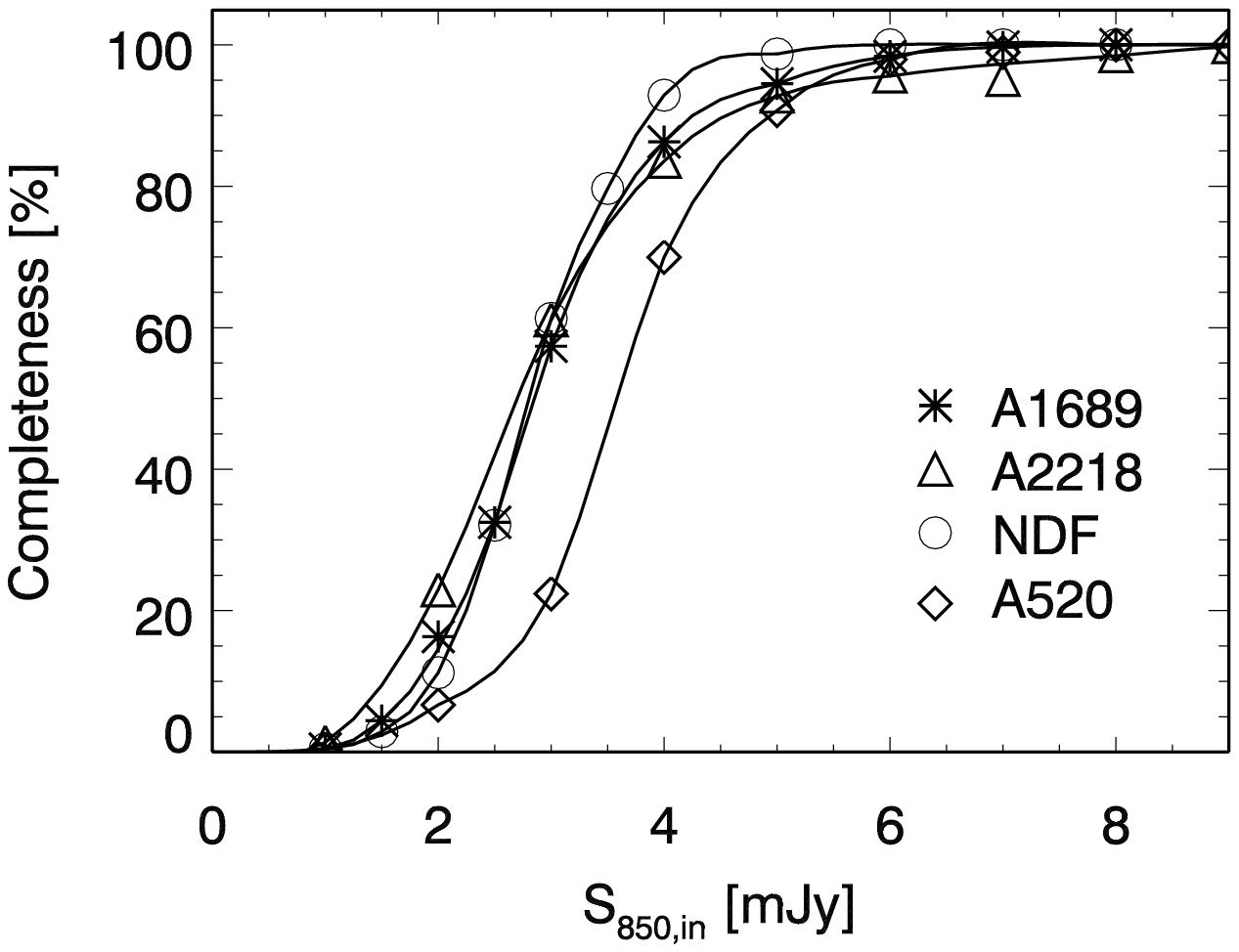}}
\center{\includegraphics[width=7cm]{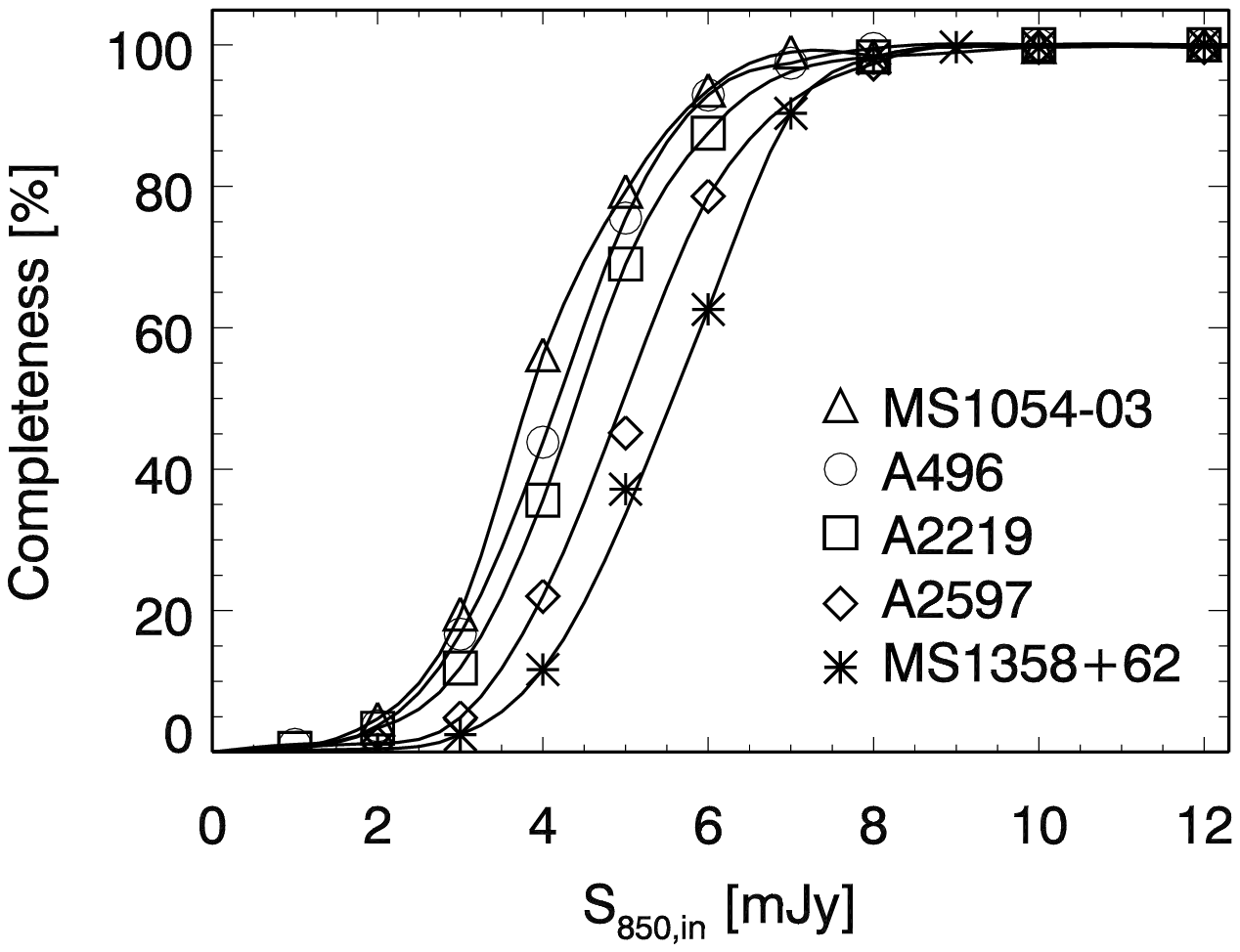}}
\center{\includegraphics[width=7cm]{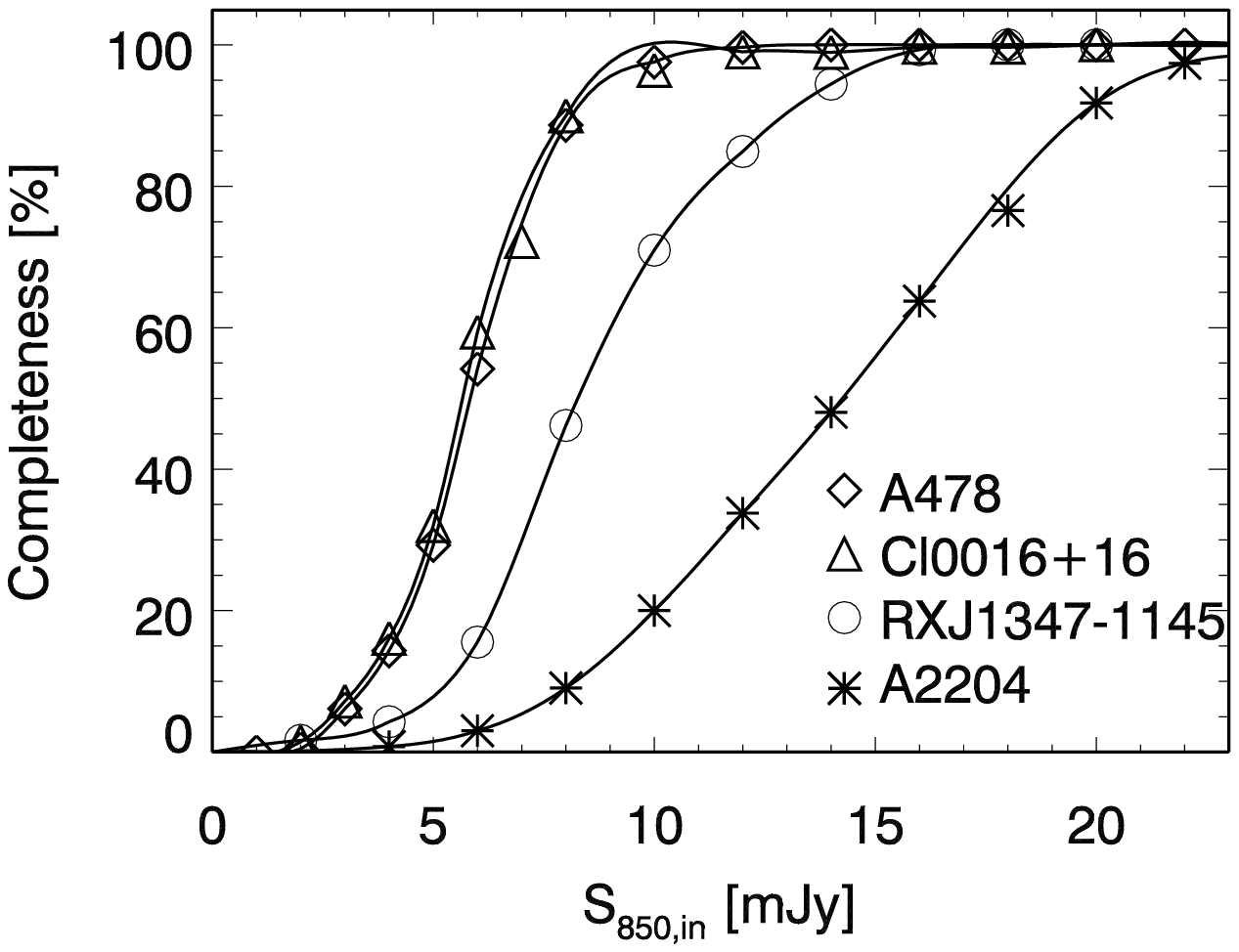}}
\caption[Completeness as function of flux]{
This figure shows the completeness as function of flux for all the 
surveyed fields.   
For a better overview the thirteen fields have been plotted in the 
three panels.  The top panel shows the deepest of the fields, while 
the middle panel shows the medium-deep fields, and the lower panel 
shows two medium-deep fields, and shallower fields.  
Note that the x-axis is different for each panel. 
\label{fig:complt}}
\end{figure}

The completeness was determined through a similar set of simulations 
as those mentioned above, where point sources are added on to empty 
maps and recovered using the MHW algorithm with the same constraints 
as for the real data.  This is done for each field, for a representative 
range of fluxes in steps of 1-2 mJy, 
repeated 4000 times for each flux level.  The 
positions were chosen to be random with a uniform distribution. 
As the simulations are performed on the whole field, which has a 
non-uniform sensitivity, the results we find is an average across each field. 
In Figure~\ref{fig:complt} we plot the completeness for the individual 
fields.  
As is seen in Figure~\ref{fig:complt}, the completeness depends on the 
depth of the observations.  For the deepest fields like A1689, A2218 and 
NTT Deep Field the observations are 80 per cent complete at a flux level of 
$\sim$\,3.5-4\,mJy and 50 per cent at $\sim$\,2.6-2.8\,mJy.  
For the other, less deep fields the 80 per cent completeness is 4.5-7.5\,mJy
and 50 per cent at 3.5-5.5\,mJy.  For A2204, which is a shallow field, 
the observations are 80 per cent complete at 18\,mJy and 50 per cent 
at 14\,mJy.

\subsection{Spurious detections}
\label{subsec:spurious}

We have addressed the issue of spurious sources.  Source detection was 
performed on inverted maps to check if any negative sources were detected. 
We ignore negative sources sitting on a chop throw, which are a 
known artifacts from the original beam pattern. 
Furthermore, we ignore the deepest fields, where confusion plays a role.   
For those fields even a detected negative 
source can be a real structure in the background.
Along the edges many negative sources were detected.  
As this indicates that many positive spurious sources would be found 
along the edge, we have decided to trim away the edge at a width of one and 
a half beam, i.e.~23$''$.  
In total we find five negative sources, which is in agreement 
with Gaussian statistics. 
Our catalogue thus may contain 5 spurious sources. 

Additionally, we have performed source extraction from the Monte Carlo 
maps (see above) for all the fields.  We find that in a hundred 
maps typically two sources were detected with $S/N\approx $ 3, and no 
sources with $S/N> $ 3.5.  This demonstrates the reliability of MHW to 
pick out real sources with $S/N > $ 3. 
With essentially no sources in the pure noise 
maps, the spurious sources detected in the inverted maps are probably 
due to the structured background.  

\citet{coppin06} used a Bayesian approach to flag potential spurious 
soures. 
In the method, which is described in detail in \citet{coppin05}, 
an {\it a priori probability distribution} is folded with the detection 
to calculate a {\it posterior probability distribution}, $P(S_j)$, for each 
individual source.   
We adopt this approach and 
the prior is found determined from simulations 
of the submm sky assuming the number counts from \citep{coppin06} and 
using the beam-pattern as known for the individual fields.  
We flag the sources, which have a more than 
5 per cent of their posterior probability distribution below 0\,mJy, 
$P(S_j<0\,{\rm mJy}) > 5$ per cent. 
These sources will be marked in the catalogue table 
(Table \ref{tab:catalogue}). 
The total number of sources is 11.  The sources are not removed from 
the catalogue, as this is only a statistical approach and does not 
allow us to discriminate unambigiously between real and spurious sources 
as is evidenced in the field MS1054-03.

\subsection{The 450\,$\mu$m maps}
\label{subsec:450maps}

The atmospheric optical depth at 450\,$\mu$m is five times larger than that 
at 850\,$\mu$m.  This makes it difficult to reliably calibrate
the 450\,$\mu$m maps, as the data are much more sensitive to any 
variation in the sky opacity.  As opposed to the 850\,$\mu$m beam 
pattern, the 450\,$\mu$m beam pattern is not well-described by a 
2-dimensional Gaussian. 
In addition, the 450\,$\mu$m beam pattern is very sensitive to any 
deformation of the JCMT dish.  Such deformations are normally the result 
of temperature variations.  Consequently, the beam pattern changes  
during the night, with changing observing conditions.  
These effects complicate a reliable source extraction from the 450\,$\mu$m.  
Furthermore, the 450\,$\mu$m beam is much narrower than the 
850\,$\mu$m beam, and more of the calibration sources may appear 
extended.  This has to be taken into account as well and might add to 
the uncertainty of the 450\,$\mu$m flux calibration.  

We have chosen not to perform deconvolution and MHW source extraction 
from the 450\,$\mu$m maps to make a separate 450\,$\mu$m 
point source catalogue.  Instead we use the maps to determine the 
450\,$\mu$m flux at the 850\,$\mu$m source positions.  That is 
done in the following way.  The $S/N$ map, the flux and the noisemap are 
used.  In a circle with radius 4$''$ around the 850\,$\mu$m source 
position, the maximum $S/N$ value is found.  If this value is equal to 
or greater than 3, the flux value in that pixel is adopted as the $S_{450}$ 
of the 850\,$\mu$m source, and the uncertainty, $\sigma_{S_{450}}$, 
is 30 per cent of that value.  If no pixels in that circle fullfil the $S/N$ 
criteria, an upper limit is given as three times the mean flux value 
corresponding to the 
three lowest noise values within the circle.   This has been done three 
times for all the 450\,$\mu$m maps, namely when the maps have been 
smoothed with a Gaussian with FWHM of 5$''$, 10$''$ and 12.7$''$.  
Smoothing with e.g.\ a 5$''$ Gaussian reduces the high spatial frequency 
noise, while smoothing with a 12.7$''$ Gaussian makes the 450\,$\mu$m 
beam equivalent to the 850\,$\mu$m beam. 
Furthermore, the smoothing reduces the effect that sources might be 
slightly extended in the original map due to intrinsic size and 
pointing errors.  
In a couple of cases the $S/N$ criterion is met with one smoothing, but 
not quite in the other one(s).  
For the reasons described above, the 450\,$\mu$m fluxes 
listed in Table~\ref{tab:catalogue} should used with caution.

%%%%%%%%%%%%%%%%%%%%%%%%%%%%%%%%%%%%%%%%%%%%%%%%%%%%%%%%%%%%%%%%%%
%%%    CONFUSION
%%%%%%%%%%%%%%%%%%%%%%%%%%%%%%%%%%%%%%%%%%%%%%%%%%%%%%%%%%%%%%%%%%

\section{Confusion}
\label{sec:confusion}

\subsection{Confusion limit in blank fields}
\label{subsec:conf_bl}

In the background of deep SCUBA maps the instrumental noise and the 
confusion noise from a fainter submm population are of approximately 
equal magnitude.  
We adopt the formalism of \citet{condon74} 
and use the same definition for the beam as \citet{hogg01}: 
$\Omega_{beam} = \pi (\theta_{FWHM}/2.35)^2$, where $\theta_{FWHM}$ is 
the width of the beam (SCUBA at 850\,$\mu$m: $\theta_{FWHM} = $ 15$''$).   
We adopt the rule of thumb that the number of sources per beam should 
not exceed one source per 30 beams before the image is considered confused 
\citep[e.g.,][]{hogg01}. 
The confusion limit is the flux, $S_{conf}$, at which the number of 
sources in the map is one source per 30 beams.  
To estimate the confusion limit, the integrated number counts, $N(>\!S)$, 
where $N(>\!S)$ denotes the number density on the sky of sources brighter 
than $S$, are used.  
For blank fields a single power law suffices to describe the 
number counts, $N(>S) = N_0 S^{-\alpha}$.  For $N(>S) = 1 / 30\Omega_{beam}$, 
the confusion limit is $S_{conf} = \sqrt[\alpha]{30\Omega_{beam} N_0}$. 
If we assume $\alpha=$ 2.0 and $N_0 = $ 13000\,deg$^{-2}$ 
\citep[based on e.g.,][]{barger99,borys03}, 
for SCUBA 850\,$\mu$m blank field observations, the confusion limit
is $\sim$\,2\,mJy.  
For an average-sized, trimmed map in our survey of 5\,arcmin$^2$, 
this number of sources is $n_s = $ 4.7.  This is close to the average 
number of sources per field in our survey. 

Confusion in the maps affects the position and flux determination.  
\citet{hogg01} have made general simulations addressing the issue of the 
errors caused by confusion.  Based on Figure 4 of his paper \citep{hogg01}, 
which gives the position error as function of detected source density where 
no prior knowledge is used in the source detection, in the Euclidean case, 
a detected source density of one per 30 beams causes a median position 
error of 0.25 times the half width at half maximum (HWHM).  At 850\,$\mu$m 
this corresponds to an additional error in the position of 1.9$''$.  For 
crowded fields in our survey where the source density is larger
than one per 30 beams, the error in the position caused by the confusion
is 4-5$''$. 
\citet{eales00} found in their simulations that in confusion-limited 
fields 10 per cent to 20 per cent of the detected
sources would lie outside an error circle of 6$''$.  Furthermore, they found
that the fluxes of the sources were boosted by a median factor of  1.44, 
albeit with a large scatter.  
However, as argued by \citet{blain01}, confusion effects will only appear 
in SCUBA maps with detection limits of 2\,mJy or less at 850\,$\mu$m; 
hence most of our data are relatively unaffected by flux boosting, 
though flux boosting is expected to play a role in the deepest maps, 
i.e.\ A1689, A2218, NDF and possibly A520. 
For these fields we adopt the Bayesian approach as used by 
\citet{coppin05,coppin06}, and as already mentioned in 
Subsect.~\ref{subsec:spurious}, to estimate how large the flux boosting 
might be.  The estimated de-boosted flux is listed in the catalogue 
Table~\ref{tab:catalogue}.  
The deboosted fluxes are about 1\,mJy fainter, though the actual number 
is connected with the $S/N$ of the detection.  

\subsection{Confusion limit in lensed fields}
\label{subsec:conf_srcpl}

For the cluster fields the confusion limit is affected by the 
gravitational lensing. 
The gravitational lensing magnifies the region seen behind the 
cluster, hence the source plane is smaller than the image plane. 
The number of beams is conserved between the image plane and the 
source plane, i.e.\ the size of the beam scales with the magnification. 
This is why it is at all possible to observe the fainter sources, which 
have a higher surface density than the brighter sources.  

The number counts in the lensed case can be written as 
$N_{lens}= (N_0 / \mu) (S / \mu)^{-\alpha}$
$= N_{blank} \mu^{\alpha-1}$, 
where $\mu$ is the gravitational lensing magnification.  
The confusion limit in the lensed case can thus be written as 
$S_{conf} = \sqrt[\alpha]{30\Omega_{beam} N_0 \mu^{1-\alpha}}$. 
As the lensing magnification varies across the field, 
We use the average magnification for a field as estimated by the ratio 
of the area in the image plane and the area in the source plane. 
This simple estimate gives an average across the field, 
which in some cases means that the estimated lensed confusion limit 
does not reflect the confusion limit of the highly magnified region 
close to the caustics. 
For the most extreme case in our sample, A1689, where the source plane 
area surveyed is 20 times smaller than the SCUBA field of view, 
the confusion limit is reduced by a factor 4.5, i.e., 
$S_{conf} = $ 0.44\,mJy. 
The confusion limits for the cluster fields based on this simple 
calculation are given in Table~\ref{tab:fields_lens}. 
In the simplified estimate of the confusion limit presented here
we assumed that the number counts are described by a 
single power law.  There are good indications that the number counts 
are described by a double power law or another function with a (gradual) 
turn-over (see Sect.\ \ref{sec:counts}).  
Including this in such a calculation will work in a favourable
direction and the confusion limit in the source plane will be lower. 

%%%%%%%%%%%%%%%%%%%%%%%%%%%%%%%%%%%%%%%%%%%%%%%%%%%%%%%%%%%%%%%%%%
%%%    GRAVITATIONAL LENSING
%%%%%%%%%%%%%%%%%%%%%%%%%%%%%%%%%%%%%%%%%%%%%%%%%%%%%%%%%%%%%%%%%%

\section{Gravitational lensing}
\label{sec:lensing}

To quantify the gravitational lensing effect we use LENSTOOL \citep{lenstool}.  
\begin{figure}%[t!]
\center{\includegraphics[width=7cm]{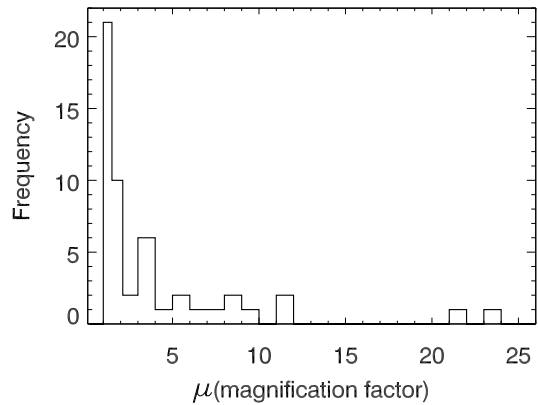}}
\caption{Histogram showing the distribution of the magnification factors. 
The histogram is binned with $\Delta\mu= $ 1, except between 1 and 2, where 
the bin-size is 0.5.  
\label{fig:hist_mag}}
\end{figure}
The gravitational potential of each galaxy cluster 
is mapped in a mass model, which describes the distribution of 
the overall potential of the cluster and to some extent of the individual 
galaxies.  For clusters where the cluster lensing is a less strong effect 
only the brightest cluster galaxies are considered in the mass model 
in addition to the global cluster potential. 
For clusters where the cluster lensing is a dominant effect many galaxies have 
been included to map the substructure of the total potential, 
as individual galaxies might cause extra lensing of the background sources.  
The lensing correction is done for the individual submm sources and 
for the sensitivity maps. 
The latter are needed in the calculation of the number counts. 
\begin{figure}%[t!]
\center{\includegraphics[width=7cm]{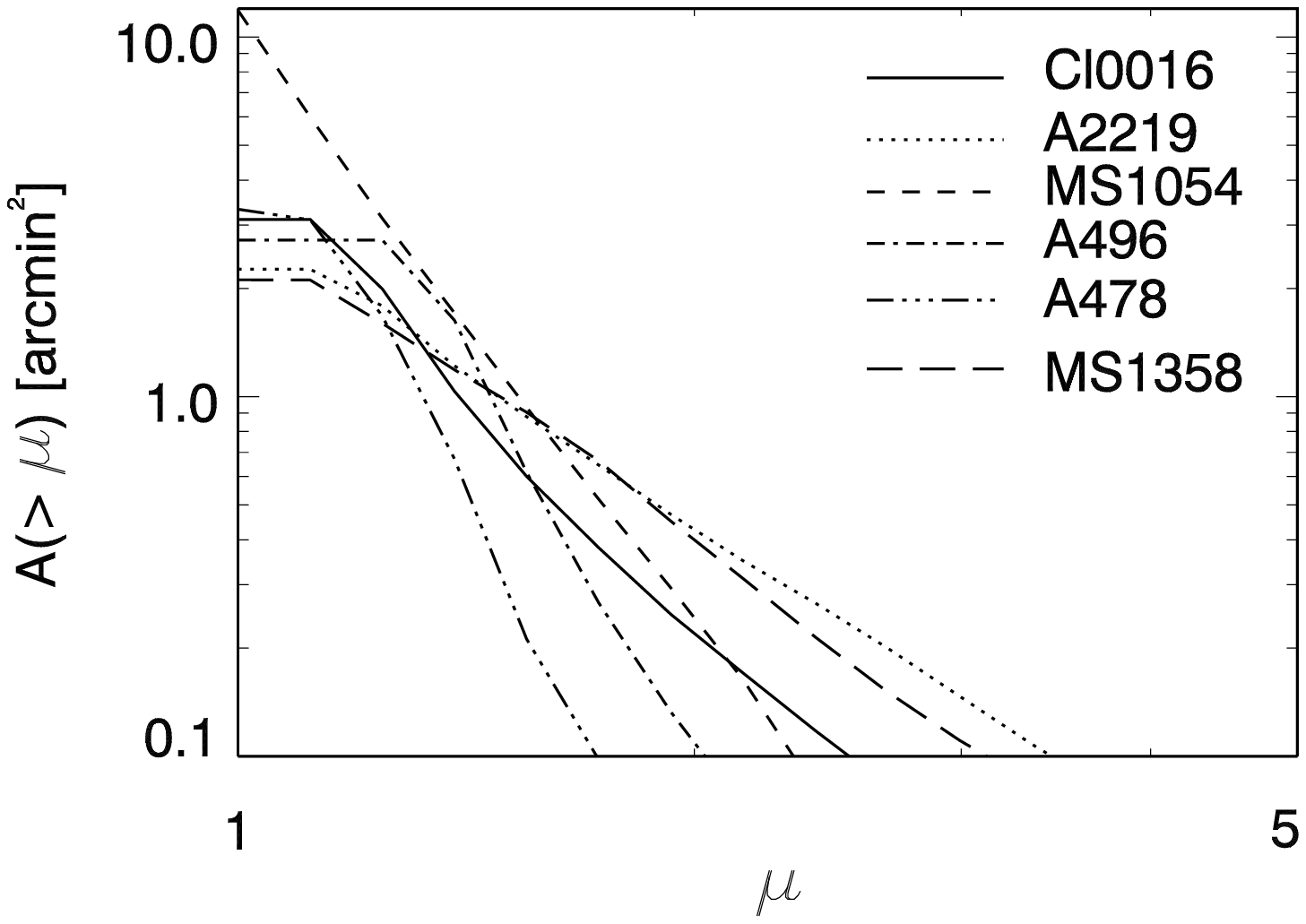}}
\vskip2mm
\center{\includegraphics[width=7cm]{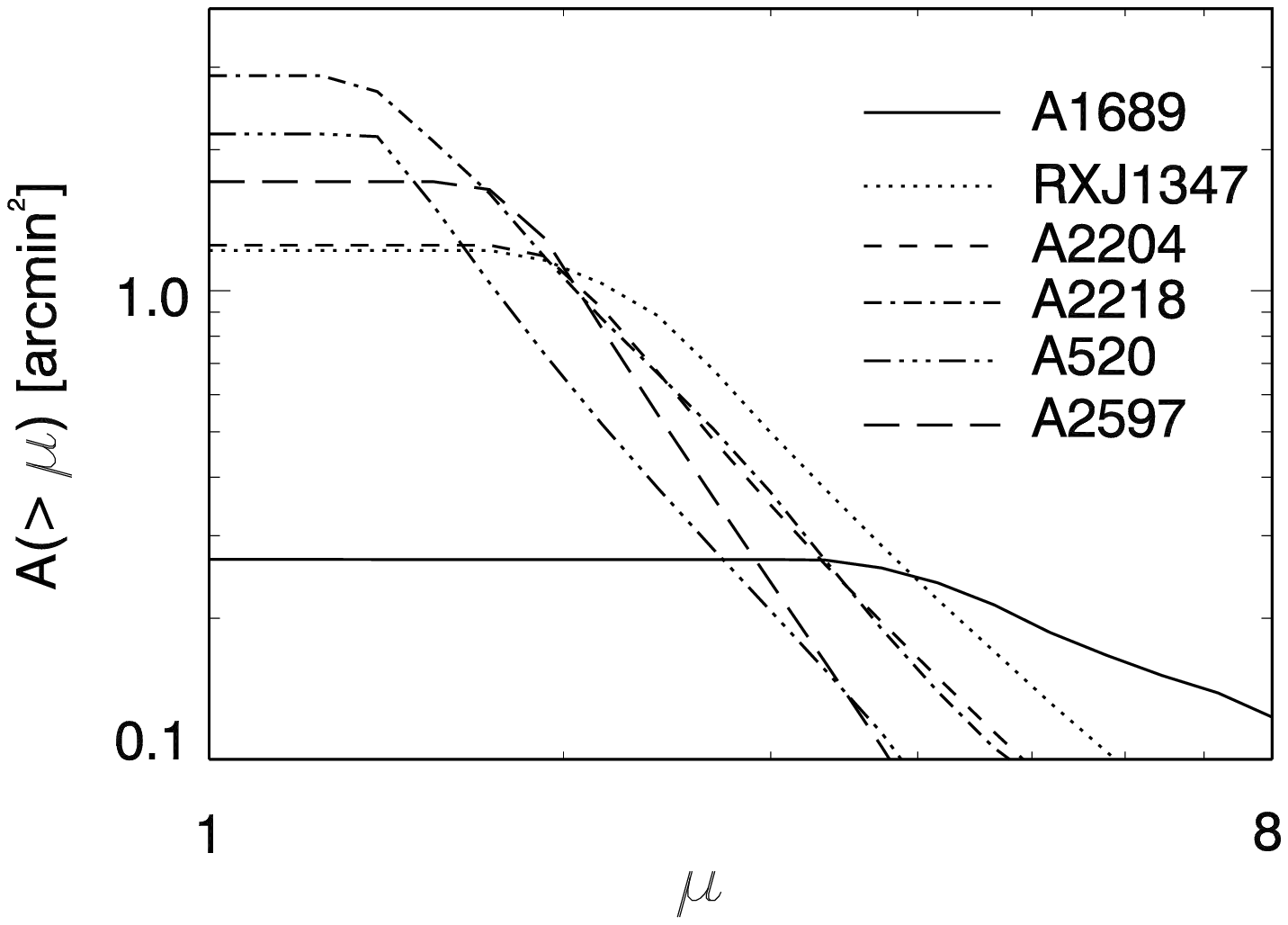}}
\caption{
The area as function of magnification for the individual fields.  
This has been split into two figures for clarity.   The top panel shows 
the fields with a relatively less strong lensing effect, while the bottom 
panel shows those with a stronger lensing effect.  We have assumed a source 
redshift of $z= $ 2.5.  Placing the source plane at a different redshift, 
$z> $ 1, would not make a significant difference.   
\label{fig:lens_area_mag}
}
\end{figure}

As the redshift is not yet known for a majority of the objects, we assume
$z= 2.5$ for the objects with unknown redshift based on the redshift 
distribution from \citet{chapman03,chapman05}. 
Likewise, the sensitivity maps, which give us the observational 
sensitivity in the image plane, are traced to a source plane at $z= $ 2.5. 
The actual redshift distribution of the faint SMGs is not known and it 
is also not known whether it follows that of the brighter SMGs as deduced
by \citet{chapman05}.  Based on a stacking analysis, \citet{wang06} 
suggest that the redshift distribution of faint SMGs peaks at redshifts 
of one or less.  However, in \citet{knudsen05} submm stacking 
results of high redshift red galaxies show that half of the EBL light 
produced by at the faint end originates from red galaxies in the redshift 
interval 1-4.  Of the five $f_{850} < 1$\,mJy SMGs with reliable 
identification and spectroscopic redshifts, the redshifts are z = 1.0, 
2.5, 2.5, 2.6 and 2.9 
\citep[][and see appendix for A1689]{Kneib_a2218,borys04,knudsen06}, 
showing no evidence for a radically different distribution than the 
one deduced by \citet{chapman05}.
We note that not knowing the exact redshifts of the SMGs will introduce 
some uncertainty in the lensing correction, however, the magnification 
correction is only weakly dependent on redshift at $z>1$, where we expect 
most of the sources to be.  

\begin{figure}%[t!]
\center{\includegraphics[width=7cm]{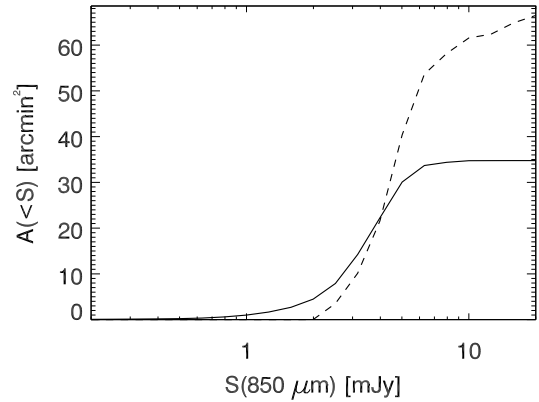}}
\caption{
The area as function of 1$\sigma$ sensitivity for the whole survey.  
The dashed line represents the area in the image plane, and 
the solid line represents the area in a source plane at redshift $z= $ 2.5. 
The actual area surveyed in the source plane is a factor two smaller than 
what we observe in the field of view of SCUBA at 850\,$\mu$m. 
For the twelve cluster fields the total area of useful data 
in the image plane is 66.5 arcmin$^2$ and in a source plane at redshift 
$z= $ 2.5 is 35\,arcmin$^2$. 
\label{fig:lens_area_sens}
}
\end{figure}

The magnification factors of the individual sources range from 1.1
to 23.  We have plotted a histogram 
of the magnification factors in Fig.~\ref{fig:hist_mag}. 
About 40 per cent of the sources are magnified by factors $\mu\,>$\,2, while 
20 per cent are magnified by 1.5\,$<\,\mu\,<$\,2, and 40 per cent have 
relatively low magnification factors 1\,$<\,\mu\,<$\,1.5.  
We have plotted the area as function of magnification factor for the 
individual fields in Figure~\ref{fig:lens_area_mag}, 
and the area as function of source plane sensitivity 
for the whole survey Figure~\ref{fig:lens_area_sens}.   
In Figure \ref{fig:lens_area_mag}, the 
two most extreme cases are A1689, where the area surveyed is a factor 
20 smaller in the source plane than in the image plane, and MS1054-03, 
where the average over the large angular area surveyed of the cluster 
dilutes the strong lensing effect caused by the core of the cluster. 
The total area observed by our survey in these 13 fields is 71.5 arcmin$^2$. 
When taking into account the gravitational magnification the area of the 
cluster fields is reduced to 35 arcmin$^2$ in the source plane. 
The area of the individual fields as well as the sensitivity in the 
source plane is listed Table~\ref{tab:fields_lens}. 
For comparison the seven cluster fields from the UK-SCUBA Lens Survey
are 40 arcmin$^2$ in the image plane and 15 arcmin$^2$ in the source 
plane \citep{sibk02}.   Likewise, in the deep though 
small survey by \citep{cowie02}, the area in the image plane 
is 18 arcmin$^2$;  assuming a reasonable amplification this corresponds 
to 6 arcmin$^2$ in the source plane.  

The uncertainties of the corrected fluxes and positions 
introduced by the lensing are in most cases small.  
The magnification is generally a monotonic function of the redshift 
(except in the very central part of a strong lensing cluster), but 
for source redshifts twice larger than the lens redshift, the 
amplification is only weakly increasing with redshift. 
As essentially all the SCUBA sources are expected to be at $z > $1, 
redshift dependence in the lensing correction is only a minor effect.  
However, the position of the source relative to the cluster centre and cluster 
members can play an important role, as the magnification can vary 
from $\sim$\,2 to $\sim$\,20 for typical background sources.  
The uncertainty in the magnification (assuming a known redshift)
is directly linked to the uncertainty in the mass model. The closer the 
object is to a critical line, the higher its magnification will be 
and the larger the uncertainty on the magnification will be.
At most four of the SCUBA sources lie relatively close to critical lines and 
therefore their magnification factors are subject to larger uncertainties.  
However, on an ensemble basis, when e.g.\ deriving the counts in terms of 
unlensed flux, the error in magnification is compensated by the error
in lensed area, thus the change in the unlensed counts due to uncertainty 
in the mass model is of the second order. 

We estimate the uncertainty of the magnification factors of the individual 
sources through a Monte Carlo simulation.  The magnification is determined 
at 1600 positions with a normal distribution within the $1\sigma$ error 
circle centered on the MHW position. 
In Table~\ref{tab:catalogue} we give the median magnification 
from the MC simulations together with the 68 per cent
deviation of the magnifications determined at the MC positions.  
For very large magnification factors, typically $> 10$, such as seen for 
the multiple-imaged galaxies in A1689, which are close to the caustics, the 
MC magnification factor distribution has both a large skewness and kurtosis. 
While we have made an estimate of the strength of possible flux-boosting, 
the results of these MC simulations show that the uncertainties on the 
magnification factor often exceeds the flux-boosting. 

Identification of multiply-imaged galaxies in the sample is important, 
as a repeated counting of the same source will affect the number counts.  
In total three multiply-imaged sources are found in the fields of 
the strongly lensing clusters A1689 and A2218.  The galaxy in A2218 
is triple-imaged with a total magnification factor of 45 and has 
been studied extensively \citep{Kneib_a2218,sheth04,kneib05,garrett05}.  
The two other galaxies are present in A1689, one is triple-imaged and 
the other is a quintuple-imaged galaxy,  both with spectroscopic 
redshift $\sim 2.5$.  Their identification will be discussed in a 
future paper.   
\begin{table}%[ht!]
\begin{center}
\caption[Parameters of the observed fields after correcting for the lensing]{
The area in the source plane at redshift $z= $ 2.5, an 
estimate of the source plane confusion limit (also see subsection 
\ref{subsec:conf_srcpl}), and the area-weighted 
1$\sigma$ sensitivity in the source plane. 
\label{tab:fields_lens}
}
\begin{tabular}{l@{\hspace{-1mm}}cccc}
\hline
\hline \\ [-2.3ex]
{\sc Cluster} & $\Omega_{s.pl.}$ & $S_{conf}$ & $\sigma_{wghtd}^{850}$  \\ 
& arcmin$^2$ & mJy &  mJy \\ [0.6ex]
\hline \\ [-2.3ex] 
Cl0016+16        &  3.1 & 1.7  &  1.48 \\
A478             &  3.3 & 1.7  &  1.65 \\
A496             &  2.7 & 1.6  &  0.98 \\
A520             &  2.2 & 1.4  &  0.78 \\
MS1054$-$03      & 11.9 & 1.8  &  1.17 \\
A1689            &  0.3 & 0.4  &  0.13 \\
RXJ1347.5$-$1145 &  1.2 & 1.0  &  1.06 \\
MS1358+62        &  2.0 & 1.3  &  1.17 \\
A2204            &  1.3 & 1.1  &  2.23 \\
A2218            &  2.9 & 1.2  &  0.50 \\
A2219            &  2.3 & 1.4  &  1.06 \\
A2597            &  1.7 & 1.3  &  0.78 \\
\hline
\end{tabular}
\end{center}
\flushleft{\vskip-1mm $\Omega_{s.pl}$ is the area of the source plane. \\
$S_{conf}$ is the flux confusion limit (see subsection \ref{subsec:conf_srcpl}). \\
$\sigma_{wghtd}^{850}$ is the area-weighted $1\sigma$ sensitivity in the source plane. }
\end{table}

%%%%%%%%%%%%%%%%%%%%%%%%%%%%%%%%%%%%%%%%%%%%%%%%%%%%%%%%%%%%%%%%%%
%%%    THE CATALOGUE
%%%%%%%%%%%%%%%%%%%%%%%%%%%%%%%%%%%%%%%%%%%%%%%%%%%%%%%%%%%%%%%%%%

\section{The catalogue}
\label{sec:catalogue}

In the twelve cluster fields we detect 54 sources and in the NTT Deep 
Field we detect 5 sources.  
The sources have been named according to their detection (SMM) and their 
J2000 coordinates.  
The catalogue of the extracted point sources 
is given in Table~\ref{tab:catalogue}.
After correcting for lensing multiplicity, 
we have detected 15 sources below the blank field confusion limit. 
Of these, 7 have flux densities $<$\,1\,mJy, which 
doubles the number of known sub-mJy sources 
(compare \citealt{cowie02}, \citealt{sibk02} and \citealt{borys04b}).  
A description of the individual fields can be found in Appendix 
\ref{app:fields} along with the final maps.
% --  Source table:   ( _the_ catalogue ;) ! )
%\input{knudsen_tab3}
\begin{table*}%[h!]
\begin{center}
\caption[List of the sources]{Catalogue of source positions, submm fluxes and uncertainties.  
$S$ is the flux, $S/N$ is the signal-to-noise of the detection in flux units, 
the $\sigma$'s give the uncertainties in the flux and position.  The uncertainties do not take into account possible additional uncertainties due to confusion. For the position, the additional confusion-uncertainty is $\sim$\,1.9$''$ 
(as described in Section \ref{sec:confusion}). 
The $^\star$ after the source name indicates that the source has a $P(S_j < 0\,{\rm mJy}) > 5\%$ (as described in subsection \ref{subsec:spurious}). 
$\mu$ is the lensing magnification, while $\mu_{\rm MC}$ is the lensing magnification from Monte Carlo simulations (see Section~\ref{sec:lensing}). 
$S_{850}$(deboost) is the deboosted flux (see subsection~\ref{subsec:conf_bl}). 
\label{tab:catalogue}}
\vskip1mm
\begin{tabular}{lcrrcccccccc}
\hline
\hline
{\sc name} & $\sigma_{pos}$ & $S_{850}$ & $S/N$ & $\sigma_{S_{850}}$ & $S_{450}$ & $S/N_{450}$ & $\sigma_{S_{450}}$ & $z$ & $\mu$ & $\mu_{\rm MC}$ & $S_{850}$(deboost) \\
           & $''$                & mJy       &       & mJy                & mJy       &             & mJy           &     &       &                &  mJy               \\
\hline
%Cl0016+16
{\em Cl0016+16} &&&&&&&&&&& \\
SMM\,J001828.9$+$162617$^\star$ & 4.0 & 6.5 & 3.2 & 2.2 & $<34.2$ & ... & ...  & {\em 2.5} & 1.2  & $1.2\pm0.02$ & \\
SMM\,J001829.4$+$162653$^\star$ & 4.0 & 5.8 & 3.2 & 2.0 & $<77.9$ & ... & ...  & {\em 2.5} & 1.2  & $1.2\pm0.02$ & \\
SMM\,J001834.2$+$162517 & 4.0 & 7.0 & 3.9 & 2.4 & $<56.1$ & ... & ...  & {\em 2.5} & 1.3  & $1.3\pm0.03$ & \\
SMM\,J001835.1$+$162559$^\star$  & 4.0 & 5.3 & 3.1 & 1.8 & $<34.2$ & ... & ...  & {\em 2.5} & 2.0  & $2.0^{+0.3}_{-0.2}$ & \\ [0.6ex]
%A478                                                                                                                       
{\em A478} &&&&&&&&&&& \\                                                                                                 
SMM\,J041322.9$+$102806$^\star$  & 3.8 & 6.8 & 3.3 & 2.5 & $<35.3$ & ... & ...  & {\em 2.5} & 1.2  & $1.2\pm0.02$ & \\
SMM\,J041323.4$+$102657 & 3.1 & 7.9 & 4.2 & 2.1 & $<53.9$ & ... & ...  & {\em 2.5} & 1.3  & $1.3\pm0.02$ & \\
SMM\,J041327.2$+$102743 & 2.3 &25.0 &14.4 & 2.8 & 55.4  & 5.3 & 16.6   &    2.837  & 1.3  & $1.3\pm0.03$ & \\
SMM\,J041328.7$+$102805$^\star$  & 3.8 & 9.0 & 3.8 & 3.3 & $<32.8$ & ... & ...  & {\em 2.5} & 1.3  & $1.3\pm0.02$ & \\ [0.6ex]
%A496                                                                                                                       
{\em A496} &&&&&&&&&&& \\                                                                                                 
SMM\,J043334.7$-$131526$^\star$ & 4.1 & 4.7 & 3.1 & 1.7 & $<43.6$ & ... & ...  & {\em 2.5} & 1.4  & $1.4\pm0.03$ & \\
SMM\,J043335.4$-$131454$^\star$ & 4.1 & 5.3 & 3.3 & 1.9 & $<54.7$ & ... & ...  & {\em 2.5} & 1.4  & $1.4\pm0.02$ & \\
SMM\,J043336.5$-$131547 & 3.2 & 4.8 & 4.0 & 1.7 &   51.1  & 3.0 & 15.3 &     0.03  & ...  & ... & \\
SMM\,J043337.4$-$131558 & 4.1 & 4.7 & 3.8 & 1.7 & $<42.9$ & ... & ...  &     0.03  & ...  & ... & \\
SMM\,J043337.6$-$131627 & 3.0 & 9.0 & 5.5 & 1.6 & $<40.3$ & ... & ...  & {\em 2.5} & 1.5  & $1.5\pm0.03$ & \\
SMM\,J043337.8$-$131541 & 3.0 & 7.9 & 5.7 & 1.4 & $<39.4$ & ... & ...  &     0.03  & ...  & ... & \\
SMM\,J043338.9$-$131444 & 4.1 & 4.0 & 3.1 & 1.4 & $<67.3$ & ... & ...  & {\em 2.5} & 1.4  & $1.5\pm0.04$ & \\
SMM\,J043339.4$-$131637 & 4.1 & 4.8 & 3.6 & 1.7 & $<52.1$ & ... & ...  & {\em 2.5} & 1.4  & $1.4\pm0.02$ & \\
SMM\,J043340.1$-$131533 & 3.0 & 6.4 & 5.1 & 1.1 & $<36.9$ & ... & ...  & {\em 2.5} & 1.5  & $1.5\pm0.04$ & \\ [0.6ex]
%A520                                                                                                                       
{\em A520} &&&&&&&&&&& \\                                                                                                 
SMM\,J045403.1$+$025547 & 3.3 & 4.7 & 4.1 & 1.1 & $<31.9$ & ... & ...  & {\em 2.5} & 1.5  & $1.5\pm0.03$  & $3.7\pm1.3$ \\
SMM\,J045406.2$+$025410$^\star$  & 4.2 & 3.9 & 3.1 & 1.4 & $<32.6$ & ... & ...  & {\em 2.5} & 5.5  & $5.5\pm0.4$  & $3.4\pm1.3$ \\
SMM\,J045406.7$+$025435$^\star$ & 4.2 & 4.3 & 3.1 & 1.5 & $<37.0$ & ... & ...  & {\em 2.5} & 4.5  & $4.4^{+0.6}_{-0.4}$ &  $2.3\pm1.6$ \\
SMM\,J045409.7$+$025510 & 3.3 & 6.0 & 4.4 & 1.4 &   29.0  & 2.7 & 8.7  & {\em 2.5} & 1.7  & $1.7\pm0.05$  & $4.7\pm1.5$ \\ [0.6ex]
%MS1054-03(big)                                                                                                             
{\em MS1054$-$03} &&&&&&&&&&& \\                                                                                         
SMM\,J105649.3$-$033606 & 3.3 & 5.0 & 3.6 & 1.1 & $<17.6$ & ... & ...  & {\em 2.5} & 1.1  & $1.07\pm0.003$ & \\
SMM\,J105655.8$-$033610 & 3.3 & 3.9 & 3.8 & 0.9 &   25.6  & 3.4 & 7.7  & {\em 2.5} & 1.1  & $1.11\pm0.007$ & \\
SMM\,J105656.3$-$033635 & 3.3 & 3.9 & 3.7 & 0.9 & $<21.2$ & ... & ...  & {\em 2.5} & 1.1  & $1.18\pm0.01$ & \\
SMM\,J105657.0$-$033612 & 2.8 & 4.9 & 4.8 & 0.9 &   61.7  & 3.6 & 18.5 & {\em 2.5} & 1.1  & $1.12\pm0.008$ & \\
SMM\,J105700.3$-$033513 & 3.9 & 3.5 & 3.2 & 1.1 & $<20.2$ & ... & ...  & {\em 2.5} & 1.1  & $1.05\pm0.003$ & \\
SMM\,J105700.3$-$033544 & 3.3 & 4.4 & 3.5 & 1.0 &   28.1  & 3.6 & 8.4  & {\em 2.5} & 1.1  & $1.08\pm0.004$ & \\
SMM\,J105701.8$-$033827 & 3.3 & 4.7 & 3.5 & 1.1 & $<27.9$ & ... & ...  & {\em 2.5} & 1.3  & $1.3\pm0.03$ & \\
SMM\,J105702.2$-$033604$^\star$  & 3.9 & 4.4 & 3.0 & 1.4 & $<21.8$ & ... & ...  &     2.423 & 1.1  & $1.11\pm0.007$ & \\
SMM\,J105703.7$-$033730 & 3.9 & 4.2 & 3.3 & 1.4 &   62.7  & 5.7 & 18.8 & {\em 2.5} & 1.6  & $1.6\pm0.05$ & \\ [0.6ex]
\hline
\end{tabular}
\flushright{\em Continues on the next page}
\end{center}
\end{table*}
\begin{table*}%[h]
\addtocounter{table}{-1}
\begin{center}
\caption[]{--- Continued from previous page.} 
\vskip1mm
\begin{tabular}{lcrrcccccccc}
\hline
\hline
{\sc name} & $\sigma_{pos}$ & $S_{850}$ & $S/N$ & $\sigma_{S_{850}}$ & $S_{450}$ & $S/N_{450}$ & $\sigma_{S_{450}}$ & $z$ & $\mu$ & $\mu_{\rm MC}$ & $S_{850}$(deboost) \\
           & $''$                & mJy       &       & mJy                & mJy       &             & mJy           &     &       &                &                    \\
\hline
%A1689                                                                                                                      
{\em A1689} &&&&&&&&&&& \\                                                                                               
SMM\,J131125.7$-$012117 & 3.3 & 5.0 & 3.9 & 1.6 & $<66.4$ & ... & ...  & {\em 2.5} & 3.9  & $3.9\pm0.4$ & $3.7\pm1.5$ \\
SMM\,J131128.6$-$012036 & 4.3 & 2.6 & 3.4 & 0.8 & ... & ... & ...      & {\em 2.5} & 23.6 & $17.6^{+28.4}_{-10.6}$ & $1.9\pm0.9$ \\
SMM\,J131128.8$-$012138 & 4.3 & 3.6 & 3.3 & 1.2 & $<32.9$ & ... & ...  & {\em 2.5} & 5.8  & $6.5^{+3.9}_{-1.0}$ & $2.4\pm1.1$ \\
SMM\,J131129.1$-$012049 & 2.8 & 4.7 & 6.0 & 0.8 & 21.4 & 4.4 & 6.4     & {\em 2.5} & 21.6 & $18.7^{+19.1}_{-6.6}$ & $4.3\pm0.8$ \\
SMM\,J131129.8$-$012037 & 4.3 & 2.5 & 3.2 & 0.8 & $<11.3$ & ... & ...  & {\em 2.5} & 3.3  & $3.1\pm1.3$ & $1.7\pm0.9$ \\
SMM\,J131132.0$-$011955 & 3.3 & 3.3 & 3.6 & 1.0 & $<14.8$ & ... & ...  & {\em 2.5} & 9.7  & $8.6^{+6.9}_{-1.5}$ & $2.4\pm1.1$ \\
SMM\,J131134.1$-$012021 & 3.3 & 3.2 & 4.0 & 1.0 & $<12.4$ & ... & ...  & {\em 2.5} & 6.5  & $6.5^{+1.1}_{-0.8}$ & $2.6\pm0.9$ \\
SMM\,J131135.1$-$012018 & 3.3 & 4.9 & 4.2 & 1.6 & $<17.6$ & ... & ...  & {\em 2.5} & 3.8  & $3.9\pm0.5$ & $3.9\pm1.3$ \\ [0.6ex]
%RXJ1347.5-1145                                                                                                             
{\em RX\,J1347.5$-$1145} &&&&&&&&&&& \\                                                                                  
SMM\,J134728.0$-$114556 & 3.0 &15.5 & 5.7 & 3.1 & 98.7 & 12.8 & 29.6   & {\em 2.5} & 3.0  & $3.0\pm0.3$ & \\ [0.6ex]
%MS1358+62                                                                                                                  
{\em MS1358+62} &&&&&&&&&&& \\                                                                                           
SMM\,J135957.1$+$623114 & 3.2 & 6.7 & 4.4 & 1.3 & $<25.5$ & ... & ...  & {\em 2.5} & 1.5  & $1.5\pm0.08$ & \\ [0.6ex]
%A2204
{\em A2204} &&&&&&&&&&& \\
SMM\,J163244.7$+$053452 & 3.2 &22.2 & 4.9 & 5.7 & $<219.5$ & ... & ...  & {\em 2.5} & 3.4 & $3.4^{+0.5}_{-0.4}$ & \\ [0.6ex]
%A2218                                                       
{\em A2218} &&&&&&&&&&& \\                                  
SMM\,J163541.2$+$661144 & 2.6 &10.4 & 7.5 & 1.4 & 53.4 & 3.5 & 16.0   & {\em 2.5} & 1.7  & $1.7\pm0.07$ & $9.5\pm1.4$ \\
SMM\,J163550.9$+$661207 & 2.4 & 8.7 &11.5 & 1.1 & 22.9 & 5.9 & 6.9    &     2.515 & 9.0  &  & $8.4\pm0.8$ \\
SMM\,J163554.2$+$661225 & 2.3 &16.1 &21.7 & 1.6 & 46.4 & 12.4 & 13.9  &     2.515 & 22   &  & $15.9\pm0.7$ \\
SMM\,J163555.2$+$661238 & 2.2 &12.8 &16.9 & 1.5 & 31.8 & 8.3 & 9.5    &     2.515 & 14   &  & $12.5\pm0.8$ \\
SMM\,J163555.2$+$661150 & 3.3 & 3.1 & 3.8 & 0.7 & 17.1 & 4.7 & 5.1    &     1.034 & 7.6  & $7.1^{+13.0}_{-2.8}$ & $2.4\pm0.9$ \\
SMM\,J163555.5$+$661300 & 2.2 &11.3 &15.8 & 1.3 & $<11.8$ & ... & ... & {\em 2.5} & 3.4  & $4.2^{+1.0}_{-0.8}$ & $11.1\pm0.7$ \\
SMM\,J163602.6$+$661255 & 3.3 & 2.8 & 3.5 & 0.6 & $<14.5$ & ... & ... & {\em 2.5} & 1.8  & $1.8\pm0.08$ & $2.1\pm0.9$ \\
SMM\,J163605.6$+$661259 & 3.1 & 5.2 & 4.9 & 0.9 & $<17.4$ & ... & ... & {\em 2.5} & 1.5  & $1.5\pm0.03$ & $4.4\pm1.1$ \\
SMM\,J163606.5$+$661234 & 3.1 & 4.8 & 4.6 & 0.8 & $<17.4$ & ... & ... & {\em 2.5} & 1.6  & $1.7\pm0.01$ & $4.0\pm1.1$ \\ [0.6ex]
%A2219                                                                                                                    
{\em A2219} &&&&&&&&&&& \\                                                                                              
SMM\,J164019.5$+$464358 & 2.9 &10.0 & 5.8 & 2.0 & 53.4 & 5.0 & 16.0   & {\em 2.5} & 1.2  & $1.2\pm0.02$ & \\ 
SMM\,J164025.5$+$464255$^\star$ & 3.7 & 5.1 & 3.1 & 1.7 & 29.4 & 2.9 & 8.8    & {\em 2.5} & 1.5  & $1.5\pm0.1$ & \\ [0.6ex]
%A2597                                                                                                                         
{\em A2597} &&&&&&&&&&& \\                                                                                                    
SMM\,J232519.8$-$120727 & 2.6 &12.3 & 7.1 & 1.8 & $<37.9$ & ... & ... &      0.08 & ...  & ... & \\
SMM\,J232523.4$-$120745 & 4.1 & 5.2 & 3.2 & 1.7 &   71.2  & 5.0 & 21.4& {\em 2.5} & 2.1  & $2.1\pm0.1$ & \\ [1.0ex]
%NTT Deep Field                                              
{\em NTT Deep Field} &&&&&&&&&&& \\                         
SMM\,J120519.0$-$074409 & 3.3 & 3.8 & 4.0 & 0.9 & $<16.0$ & ... & ... &           &   &  & $3.0\pm1.1$ \\
SMM\,J120520.6$-$074448 & 4.0 & 3.0 & 3.2 & 0.8 & $<13.7$ & ... & ... &           &   &  & $2.0\pm1.1$ \\
SMM\,J120522.1$-$074431 & 3.3 & 3.5 & 3.9 & 0.9 & $<14.0$ & ... & ... &           &   &  & $2.7\pm1.0$ \\
SMM\,J120523.1$-$074516 & 4.0 & 3.4 & 3.4 & 0.9 & $<16.7$ & ... & ... &           &   &  & $2.3\pm1.2$ \\
SMM\,J120525.1$-$074512 & 3.3 & 4.0 & 4.3 & 0.9 & $<23.8$ & ... & ... &           &   &  & $3.3\pm1.0$ \\ [0.6ex]
\hline
\end{tabular}
\end{center}
\end{table*}

% ====================================================================

%%%%%%%%%%%%%%%%%%%%%%%%%%%%%%%%%%%%%%%%%%%%%%%%%%%%%%%%%%%%%%%%%%
%%%    SOURCE EXTRACTION
%%%%%%%%%%%%%%%%%%%%%%%%%%%%%%%%%%%%%%%%%%%%%%%%%%%%%%%%%%%%%%%%%%

\section{Number counts}
\label{sec:counts}

\subsection{Determining the number counts from maps of non-uniform sensitivity}

The notation $N(>S)$ is typically used for cumulative number counts: 
the number of sources per unit solid angle brighter than a flux limit $S$. 
Calculating the cumulative number counts by counting the number of 
sources with $>S$ must be done on a map with uniform sensitivity.  
SCUBA maps, however, do not have uniform sensitivity.  
The problem of determining number counts for maps of non-uniform 
sensitivity has previously been discussed for several of the 
blank field surveys \citep{webb_3h,borys03,coppin06,scott06}. 
The presence of gravitational lensing results in even larger 
non-uniformity compared to some of the large blank field surveys and 
complicates any completeness corrections.  We here present the number 
counts as deduced using two different approaches.  
In both cases cluster members like cD galaxies are excluded and also the 
sources, which we have marked as potentially spurious using the scheme 
from \citet{coppin05} (see subsection~\ref{subsec:spurious}).  
Taking into account that the multiply-imaged sources each count as one, 
the total number sources in the number counts analysis is 40.
\begin{figure*}
\begin{center}
\center{\includegraphics[width=11cm]{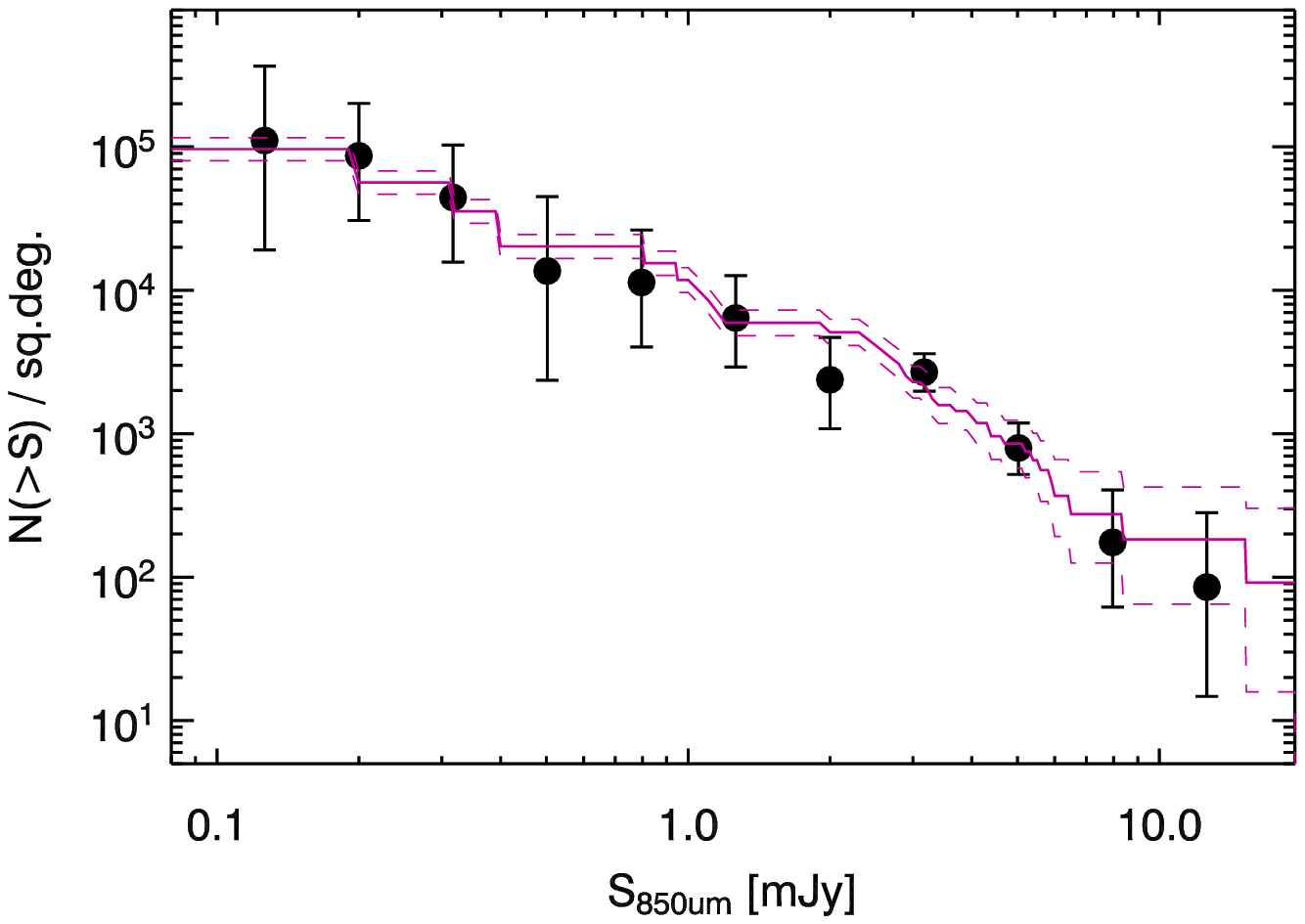}}
\center{\includegraphics[width=11cm]{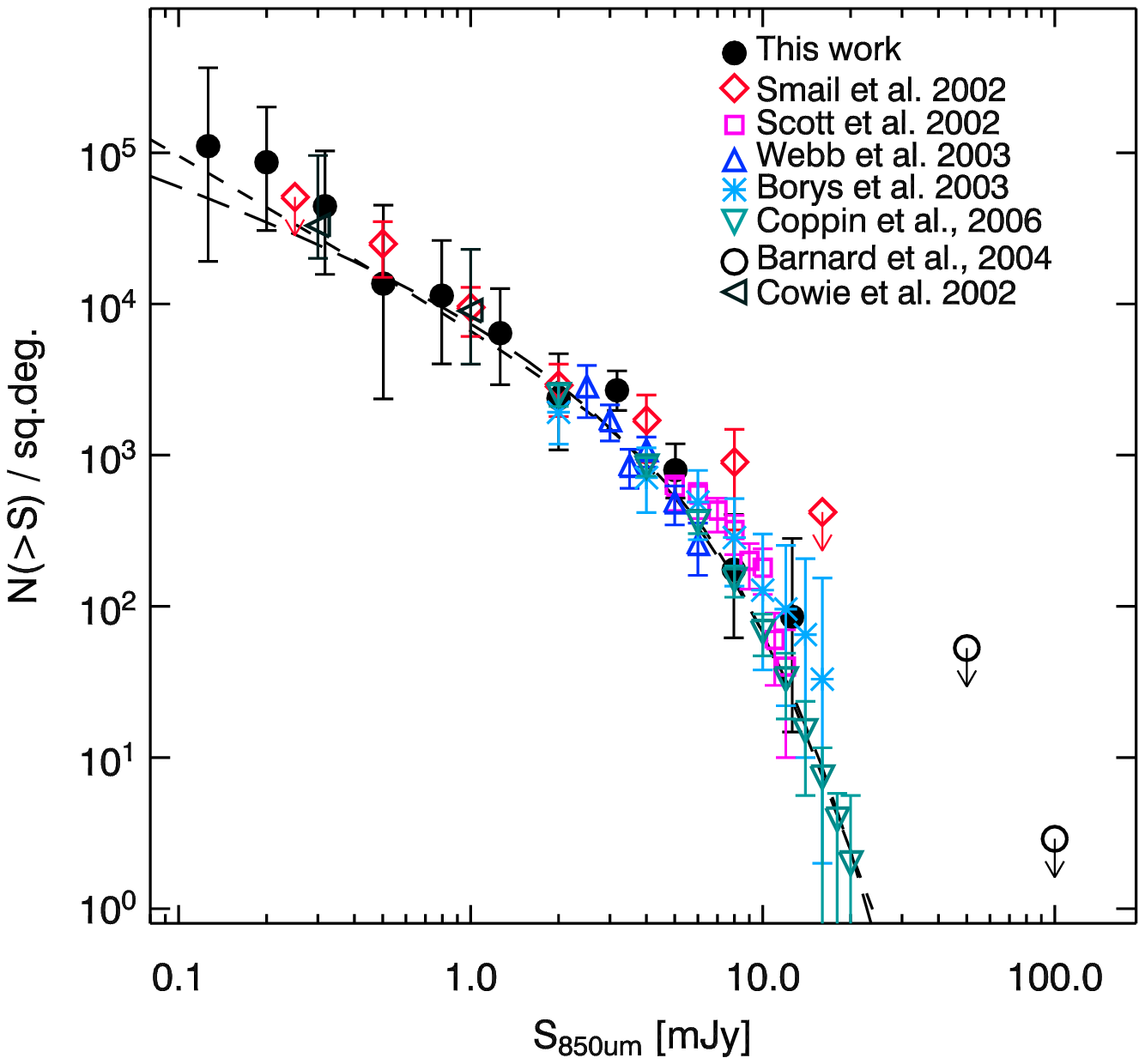}}
\caption[The 850\,$\mu$m number counts]{
The 850\,$\mu$m number counts, $N(>\,S)$, as determined from this work.   
The number counts are determined based on the data from 12 cluster fields 
and one blank field.  The data from the cluster fields have first been 
corrected for the gravitational lensing.   
Seven sources have been detected with sub-mJy fluxes.  This doubles the 
number of such sources known \citep[compare][]{cowie02}.  
{\em Upper panel:} \ 
The black dots show the results from the first approach for determining the 
number counts and the solid (magenta) line bracketed by the dashed show the 
results of the second approach (for details see text). 
{\em Lower panel:}  The number counts from other 
fields have been plotted together with the number counts from this work. 
The short dashed line shows the best-fit double powerlaw, 
Eq.~\ref{eq:doubpowlaw}, and the long dashed line shows the best-fit 
Schechter function, Eq.~\ref{eq:schechter}. 
\label{fig:counts}}
\end{center}
\end{figure*}

The first approach:  
For a given flux level $S$, only the 
surveyed area where 3$\sigma <S$ is considered.  The $N(>S)$ is then 
the number of sources with $>S$ within that area divided by the source 
plane area, $\Omega_{<S}$. 
For the cluster fields we use the fluxes and sensitivity maps 
corrected for the gravitational lensing, as described in the 
previous section.   
Upper and lower errors are calculated using Poisson statistics 
weighted by the area, $\Omega_{<S}$.  
We use the tables for confidence limits on small
numbers of events from \citet{gehrels86}.  
It should be noted that at each $S$ only a small number 
of sources is counted, in particular at the faint and bright end, as 
is reflected in the error bars.  
The resulting number counts for the 850\,$\mu$m observations are 
plotted in Figure~\ref{fig:counts}.  
The number counts, $N(>S)$, the number of sources for each data point and 
the area, $\Omega_{<S}$, are given in Table~\ref{tab:counts}.  
Due to the non-uniform sensitivity across the observed fields, 
the area $\Omega_{<S}$ varies with $S$.  Consequently, $N(>S)$ is 
not uniformly decreasing with $S$, which would otherwise be 
expected for cumulative number counts determined in fields with 
uniform sensitivity.  
We note that even though the area of the circle with diameter 15$''$ is 
0.049\,arcmin$^2$, 
and that the counts for $S_{850} < $ 0.2 are calculated from an area that 
appears to be smaller than the beam in the image plane, it should be 
remembered that the area is in the source plane, which due to the 
large magnification would appear as a 
much larger area in the image plane corresponding to many beams.  

We use a second, alternative approach to estimate the number counts: 
\begin{equation}
N(>S) = \int^{\infty}_{S}  \frac{n(S')}{\Omega_{eff}}   \mathrm{d} S' \ ,
\end{equation}
where $n(S')$ is the number density of sources within the flux interval 
$S'$ and $S'+\mathrm{d}S'$.  $\Omega_{eff}$ is the effective area over 
which the survey is sensitive to sources with flux $>S$ is given by 
\begin{equation}
\Omega_{eff} = \Omega_{total} \cdot C(S) \ .
\end{equation}
$\Omega_{total}$ is the total area of the survey. 
$C(S)$ is the completeness function, that also takes into account the 
effects of the lensing, which is determined through simulations as follows.  
Sources are placed at random in the source plane and run through the 
lensing models (Sect.\ \ref{sec:lensing}) to determine their magnification. 
This is done for each flux level $S$ and compared with the sensitivity 
map to determine how large a fraction of the simulated sources would be 
detectable.  
For the sources from the deepest fields we use their deboosted fluxes. 
Similarly to above, the upper and lower errors are calculated using Poisson 
statistics weighted by the area and use the tables for confidence limits 
on small numbers of events from \citet{gehrels86}.  
This approach allows us to take into account the deboosted fluxes of 
the deepest maps.  
The results are also plotted in Figure~\ref{fig:counts}.
As opposed to the previous approach, the error bars are very small at the 
faint fluxes due the larger total number of sources.  We caution that 
these error bars do not contain systematic errors caused by 
e.g.\ uncertainties in the lensing correction.

The number counts are probed to the faintest source in the survey, 
which has a lensing corrected flux of 0.11\,mJy.   
The faint end of the number counts is dominated by the two cluster 
fields A1689 and A2218, which on the other hand do not contribute 
much at the bright end. 
A tentative analysis of the number counts show that the counts are 
not well-described by a single power-law function, but are better 
described by a double power-law 
\begin{equation}
\label{eq:doubpowlaw}
\frac{{\rm d} N(>S)}{{\rm d}S} = 
              \frac{N_0}{S_0} \bigg ( \bigg (\frac{S}{S_0}\bigg )^\alpha + 
                             \bigg (\frac{S}{S_0}\bigg )^\beta \bigg )^{-1}\ ,
\end{equation}
or another function with a turn-over, such as a Schechter function 
\begin{equation}
\label{eq:schechter}
\frac{{\rm d} N(>S)}{{\rm d}S} = \frac{N_0}{S_0} S \bigg ( \frac{S}{S_0}\bigg )^\alpha \exp \bigg ( \frac{-S}{S_0}\bigg ) \ .
\end{equation}
We have performed a simple $\chi^2$-analysis for these two functions, 
as these two have been used previously for numerical modeling or analysis 
of SCUBA number counts.  For this we have also included the 
number counts from the SHADES survey \citep{coppin06} 
as these provide a better constraint at the brighter end.  
We have also included the additional constraint that the integrated 
light well below 0.1 mJy should not be larger the extragalactic 
background light \citep{puget96,fixsen98}. 
We note that the slope at the faint end is diverging and if it was to 
continue to much fainter fluxes it would result in an overproduction 
of the background light. 
The resulting parameters from this analysis are given in 
Table~\ref{tab:ns_fits} as well as the best fit is overplotted on top 
of the number counts shown in Figure~\ref{fig:counts}.
\begin{table}%[t!]
\begin{center}
\caption[Number counts]{The 850\,$\mu$m number counts. 
\label{tab:counts}}
\begin{tabular}{c@{\hspace{5mm}}c@{\hspace{5mm}}c@{\hspace{5mm}}c}
\\ 
\hline
\hline \\ [-2mm]
$S_{850}$ & $N(>S)$ & N$_{src}$ & $\Omega_{<3\sigma}$ \\
mJy & arcmin$^{-2}$ &         & arcmin$^2$ \\
\hline 
& & & \\ [-1.6ex]
 0.13 & $30.8  _{-25.4}  ^{+70.7}$  &  1 & 0.033 \\ [0.2ex]
 0.20 & $24.1  _{-15.5}  ^{+31.8}$  &  2 & 0.083 \\ [0.2ex]
 0.32 & $12.3  _{-7.96}  ^{+16.3}$  &  2 & 0.16  \\ [0.2ex]
 0.50 & $3.88  _{-3.13}  ^{+8.71}$  &  1 & 0.26  \\ [0.2ex]
 0.80 & $3.15  _{-2.04}  ^{+4.16}$  &  2 & 0.63  \\ [0.2ex]
 1.26 & $1.78  _{-0.97}  ^{+1.73}$  &  3 & 1.69  \\ [0.2ex]
 2.00 & $0.66  _{-0.36}  ^{+0.64}$  &  3 & 4.54  \\ [0.2ex]
 3.17 & $0.75  _{-0.20}  ^{+0.26}$  & 14 & 18.8  \\ [0.2ex]
 5.02 & $0.22  _{-0.076} ^{+0.11}$  &  8 & 36.2  \\ [0.2ex]
 7.96 & $0.049 _{-0.031} ^{+0.064}$ &  2 & 41.2  \\ [0.2ex]
12.6  & $0.024 _{-0.020} ^{+0.054}$ &  1 & 42.2  \\ [0.2ex]
\hline
\end{tabular}
\end{center}
\end{table}
\begin{table}%[t!]
\begin{center}
\caption[Fits to number counts]{The resulting parameters from fits to the
850\,$\mu$m number counts. 
\label{tab:ns_fits}}
\begin{tabular}{c@{\hspace{2mm}}c@{\hspace{2mm}}c@{\hspace{2mm}}c@{\hspace{2mm}}c}
\hline
\hline \\ [-2mm]
Function & $N_0$ & $S_0$ & $\alpha$ & $\beta$  \\
%& mJy & arcmin$^{-2}$ &         &  & \\
\hline 
& & & \\ [-1.6ex]
 Eq.~\ref{eq:doubpowlaw} & $658\pm48 $ & $9.60^{+0.30}_{-2.12}$ & $2.12^{+0.14}_{-0.08}$ & $6.22^{+0.51}_{-0.34}$ \\ [0.2ex]
 Eq.~\ref{eq:schechter} & $1039\pm69 $ & $4.30\pm0.08$ & $-2.62\pm0.10$ & ... \\ [0.2ex]
\hline
\end{tabular}
\end{center}
\end{table}

\subsection{Fluctuation analysis}

We have performed a fluctuation analysis, or $P(D)$ analysis,  
on the NDF, A1689 and A2218 fields.  
This has previously been done for blank field (sub)mm data by 
\citet{hughes98} and \citet{maloney05} as a statistical method to 
probe the number counts fainter than the sensitivity limit of the data.  
We measure the pixel distribution from simulated maps, which were 
created using an input source distribution, convolved with the beam 
and added to the Monte Carlo maps from Section \ref{sec:src_ext}.  
As number counts for the input source distribution we used the 
Schechter function, Eq.~\ref{eq:schechter}, stepping through the three
different parameters.  The positions of the simulated sources were 
drawn from a set of random positions with a normal distribution without
taking into account clustering.  For the A1689 and the A2218, we 
calculated the effects of the gravitational lensing of the simulated 
sources, i.e.\ the magnification and position in the image plane. 
We caution that gravitational lensing will introduce similar 
uncertainties and effects as when calculating the lensing effects 
for the real sources and hence adding an extra complication for the 
interpretation of the results.   
The $P(D)$ of the simulated data is compared to that of the real data 
for the three fields to determine the parameter set of the number 
counts that best fit the real data.  The resulting parameters are listed 
in Table~\ref{tab:flucan} along with some calculated counts values for 
comparison with the number counts determined in the previous subsection.  
As seen for both the NDF and A1689, $\alpha < -3.5$, gives a very steep 
function especially at the faint end which would strongly overproduce 
the EBL.  The resulting parameters from A2218 are reasonably close to 
those deduced the Schechter function fit to the number counts.  
\begin{table*}%[t!]
\begin{center}
\caption[Results from fluctuation analysis]{Results from the fluctuation analysis.
\label{tab:flucan}}
\begin{tabular}{lcccccccc}
%\begin{tabular}{l@{\hspace{5mm}}c@{\hspace{5mm}}c@{\hspace{5mm}}c@{\hspace{5mm}}c@{\hspace{5mm}}c}
\hline
\hline \\ [-2mm]
Field & $N_0$ & $S_0$ & $\alpha$ & $N(>2$mJy$)$ & $N(>1$mJy$)$ & $N(>0.5$mJy$)$ & $N(>0.2$mJy$)$ \\
\hline
NDF  &  550 & 4.0 & -3.75 & 1750\,deg$^{-2}$ & 8700\,deg$^{-2}$ & ... & ... \\
     &      &     &       & 0.49\,arcmin$^{-2}$ & 2.4\,arcmin$^{-2}$ & ... & ... \\
A1689 & 650 & 4.0 & -3.5 & ... & ... & 24600\,deg$^{-2}$ & 114600\,deg$^{-2}$ \\
      &     &     &      & ... & ... & 6.8\,arcmin$^{-2}$ & 31.8\,arcmin$^{-2}$ \\
A2218 & 750 & 6.5 & -2.75 & ... & ... & 30700\,deg$^{-2}$ & 72400\,deg$^{-2}$\\
      &     &     &       & ... & ... & 8.5\,arcmin$^{-2}$ & 20.1\,arcmin$^{-2}$  \\
\hline 
\end{tabular}
\end{center}
\end{table*}
We note that although all fields are roughly equally deep 
in the image plane, the faint counts are best probed by A2218, since NDF 
is not gravitationally lensed so that that field does not probe faint 
fluxes very well, and A1689 covers only a very small area in the source 
plane. 
As $P(D)$ analysis is a statistical tool it is best applied on large fields 
as was done by e.g.\ \citet{maloney05}.

\subsection{Comparison with other surveys}

Here we will compare the derived number counts with those determined 
through other surveys, both lensed surveys and blank field surveys.  
The number counts from other studies have been plotted together 
with the number counts from this work in Figure~\ref{fig:counts}.

{\it Lensing surveys:}\ 
Three other studies of SCUBA observations of cluster fields have been 
published:  
For the UK-SCUBA Lens Survey seven cluster fields were targeted 
and number counts were determined to $S_{850} = $ 0.5\,mJy 
\citep[e.g.,][]{smail97,blain99a,sibk02}. 
\citet{cowie02}  obtained deeper SCUBA observations of three 
of these fields.
A shallower cluster survey was performed by \citep{chapman02}, 
in which eight clusters were observed with SCUBA, however,   
no sub-mJy sources were detected.  For $>1$\,mJy, our number counts 
agree with those of \citep{chapman02}.  Here we will focus the comparison 
on the surveys from \citet{cowie02} and \citet{sibk02}.  As both those 
surveys are relatively small in area, a comparison is only interesting 
where such surveys have their strength, namely at the faint fluxes.  
\citet{cowie02} detect five sub-mJy sources.  We detect 
seven sub-mJy sources, and thereby double the number of known sub-mJy 
sources.  Our faint number counts are in good agreement with those of 
Cowie et al.\ and Smail et al. 

{\it Blank field surveys:}\ 
Blank field surveys are surveys with no strongly gravitationally lensing 
clusters present in the surveyed area.  Such surveys typically cover 
much large areas than the lensed surveys, and are limited in depth by 
the blank field confusion limit ($\sim$\,2.0\,mJy).  
Hence the strength of those surveys lies at brighter fluxes.  
Several such surveys have been carried out:  
The Canada-UK Deep SCUBA Survey (CUDSS) \citep{eales00,webb_3h}, 
covered 75\,arcmin$^2$ to the blank field confusion limit. 
The 8mJy-survey (Scott et al.) covered an area of 260\,arcmin$^2$
to a flux limit $>$\,5\,mJy/beam.  
The Hubble Deep Field North (HDF-N) has been surveyed extensively, 
which has been brought together in the so-called ``HDF-N SCUBA supermap''
by \citet{borys03}, which covers 165\,arcmin$^2$ 
to depths between 0.4 and 6 mJy/beam. 
\citet{barger99} have surveyed the Hawaii Survey Fields 
covering an area of 104\,arcmin$^2$ to a flux limit of 8\,mJy 
with a small area of 7.7\,arcmin$^2$ almost to the confusion limit.  
\citet{coppin05} surveyed 70 arcmin$^2$ of the Groth Strip to a 
depth of $1\sigma$ rms.\ $\approx 3.5$\,mJy. 
A re-analysis of several blank field surveys was carried out by 
\citet{scott06}.  Finally the results of the SHADES survey were  
published by \citet{coppin06}, where 720\,arcmin$^2$ were covered to a 
noise level of $\sim 2$\,mJy uncovering more than a 100 submillimetre 
galaxies.  Through a detailed analysis \citet{coppin06} determine  
the first differential submm number counts and fit their results with 
a double power law.  
With minor deviations, there is an overall good agreement between the
bright end the number counts of the work presented here and previous 
work.  Though, we do find that the slope of the power-law at the bright 
end is a bit steeper than previous work ($\alpha\sim $ 1.9-2.2). 

\subsection{Resolving the extragalactic submillimetre background light}

Using the differential number counts from equation \ref{eq:doubpowlaw}, 
we calculate the integrated background light.  At $S\sim $ 0.10\,mJy, 
the integrated background produced by our sources is comparable to the 
background light detected with COBE \citep{puget96,fixsen98}.  
Given that sources with fluxes below 0.1\,mJy also contribute to the 
integrated background, there is a possibility that our counts overpredict 
the integrated background somewhat. 
The overproduction of the background light, which is caused by the shape 
of the number counts at the faint end, is possibly due to the low number 
statistics. 
Given the differential counts from equation \ref{eq:doubpowlaw} the 
dominant contribution to the integrated background light comes from the 
sources with fluxes $S_{850}$ between 0.4\,mJy and 2.5\,mJy with 50 per cent
of the background resolved at 1\,mJy.  The latter is in agreement 
with the results from \citet{sibk02}, \citet{cowie02} and \citet{chapman02}.
Sources with $S_{850} > $ 2.5\,mJy contribute $\sim$\,25 per cent to the 
integrated background, of which sources with $S_{850} > S_0$ contribute 
only $\sim$\,10 per cent. 
This means that the bulk of the submm energy output from the submm galaxy 
population arises from sources just fainter than the blank field confusion 
limit.

% ====================================================================

\section{Conclusions}
\label{sec3:conclusion}

We have conducted a deep submm survey using SCUBA.  We have observed twelve
clusters of galaxies and the NTT Deep Field.  The total area surveyed 
is 71.5\,arcmin$^2$ in the image plane.  For the cluster fields 
the total area in the source plane is 35\,arcmin$^2$.  
This is the largest deep submm lens survey of its type to date. 
The gravitational \ lensing \ reduces \ the \ confusion \ limit \ allowing for 
observations of \ sources with $S_{850} < $ 2\,mJy. 
The data have been analysed using Monte Carlo simulations to 
quantify the noise properties, Mexican Hat Wavelets have been used 
for source extraction and simulations were performed to quantify the 
error of the analysis.   
\begin{itemize}
\item In total 59 sources have been detected, 
of which 10 have flux densities below the blank field confusion 
limit.  Four are associated to cluster cD galaxies. 
Three sources in the field of A2218 are multiple images of the 
same galaxy, and in A1689 five are associated with two multiply-imaged 
galaxies.  
\item 
The number of sub-mJy sources is seven, which doubles the 
number of such sources.  
\item The integrated number counts are probed 
over two decades in flux down to 0.10\,mJy.  
The number counts cannot be described by a single power law, but have to 
be described by a double power law or another function with a turn-over.  
Describing the differential counts by a double power law function we 
find that the turn-over is $\sim$\,6\,mJy.  
At 1\,mJy the number counts are $\sim$\,10$^4$\,deg$^{-2}$ and at 
0.5\,mJy they are $\sim$\,2$\times$10$^4$\,deg$^{-2}$, based on 
derived differential counts.  
\item Another key result is that
essentially all of the integrated submm background is resolved. 
At 1\,mJy 50 per cent of the background is resolved, and at 0.4\,mJy 75 per 
cent is resolved.  The dominant contribution to the background comes from 
sources with fluxes $S_{850}$ between 0.4\,mJy and 2.5\,mJy, while the 
bright sources with fluxes $S_{850} > $ 6\,mJy contribute only 10 per cent. 
This means that the bulk of the energy comes from submm galaxies with 
fluxes just below the blank field confusion limit. 
\end{itemize}

The submm number count distribution is an observable for the submm galaxy 
population as a whole and provides strong constraints on the models 
describing the submm galaxy population and their evolution.  
While the submm number counts are well studied at the bright end 
($>2$\,mJy), the faint end ($<2$\,mJy) and the extremely bright end 
($>20$\,mJy) remain difficult to probe. 
The extremely bright end is challenged by the steep counts. 
The faint end is challenged by the blank field confusion limit.  
The present survey has made a substantial contribution to the faint 
end, however, it essential to follow this through with future 
instrumentation such as more sensitive instruments like SCUBA-2 and 
LABOCA, with which an even larger number number of strongly lensing 
clusters can be surveyed, larger telescopes such as the LMT and CCAT 
for which the blank field confusion limit will be lower, and of course 
ALMA which will be able to study the faint sources in great detail.   

% ====================================================================

\section*{Acknowledgements}
  We thank Vicki Barnard, Tracy Webb, Marijn Franx and Graham Smith 
  for fruitful discussions. 
  We are grateful to Patricio Vielva and his colleagues for useful 
  discussions regarding wavelets and letting us use their software. 
  We thank an anonymous referee for constructive comments, which helped 
  improve the manuscript. 
  The JCMT is operated by the Joint Astronomy Centre on behalf of the 
  Science and Technology Facilities Council of the United Kingdom, 
  the Netherlands Organization for Scientific Research and the National 
  Research Council of Canada.    
  KKK acknowledges support by the Netherlands Organization for Scientific 
  Research (NWO) and the Leids Kerkhoven-Bosscha Fonds for travel support. 
  JPK acknowledges support from Caltech and CNRS.

% ====================================================================

\appendix

\section{Description of the individual fields}
\label{app:fields}

% -- map-mosaic 1
\begin{figure*}%[ht!]
\begin{center}
\center{\includegraphics[width=12.9cm]{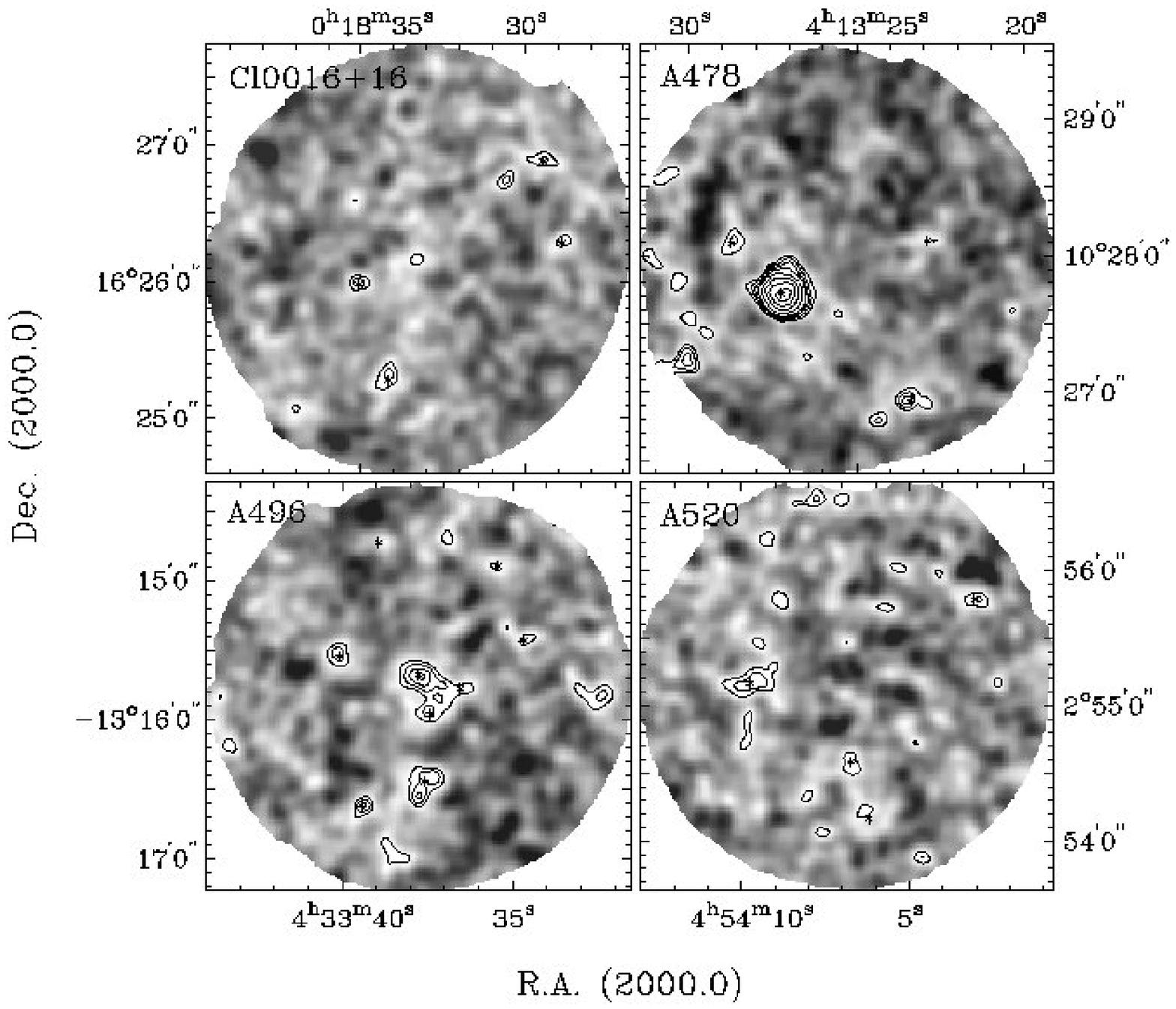}}
\caption[The maps]{
The signal-to-noise ($S/N$) 850\,$\mu$m SCUBA maps of the clusters 
Cl0016+16, A478, A496 and A520. 
The overlayed contours represent $S/N= $ 3,4,5,6,7, but for A478 the 
contours represent $S/N = $ 3,4,5,6,8,10,12,14.   The stars $\star$ 
indicate the position of the detected sources.  
\label{fig:all_maps1}}
\end{center}
\end{figure*}
%\\ \\
\underline{\it Cl0016+16}\ \  
In this field, four point sources were detected with $3 < S/N < 4$.  
The magnification factors for the four sources are between 1.2 and 2.2. 
The corrected fluxes of the sources are between 2.4 and 5.2 mJy.
We note that three sources have more than 5 per cent of their posterior 
probability distribution below 0 mJy. 
None of the sources have significant 450\,$\mu$m flux detections. 
A shallower SCUBA map of this field is presented in \citet{chapman02}, 
where the depths is about a factor two shallower than our map.  
Within the positional uncertainties the two 
sources from Chapman et al.\ are coincident with SMM\,J001834.2+162517
and SMM\,J001835.1+162559, though we find that the observed fluxes are 
fainter than in Chapman et al. 
The mass model is based on the results from Natarajan (private 
communication) with a substantially more detailed description compared 
to that of Chapman et al. 
\\ \\
\underline{\it A478}\ \  This cluster is well-known in cooling-flow studies 
\citep[e.g.,][]{white91}. 
Some of the SCUBA data presented here were obtained by others to study the 
cooling-flow, however, the cooling-flow has not been detected in the 
data.   Four point sources are detected. 
With a detected flux of $S_{850} = $ 25$\pm$3\,mJy the source 
SMM\,J041327.2+102743 is the brightest source in the survey.  This source 
has been studied in detail and is identified with a type-1 quasar
at redshift $z= $ 2.837 \citep[][there denoted SMM\,J04135+10277]{knudsen03}. 
The three fainter sources have $S/N$ between 3.3 and 4.2. 
All four sources have magnification factors of 1.2-1.3.  The fluxes 
of the three other sources are 5.6-7.3\,mJy. 
We note that two sources have more than 5 per cent of their posterior 
probability distribution below 0 mJy. 
The quasar is the only source in this field with 450\,$\mu$m flux 
detection.  
Close to the SE-edge a fifth bright source is detected, however, as 
it is less than 1.5 beam from the edge it is not included in the 
catalogue. 
The mass model is a simple model, which we constructed based on the 
published velocity dispersions:  the model includes the cluster potential 
and the potential of the cD galaxy \citep{zabludoff90,allen93}.
\\ \\
\underline{\it A496}\ \  This is the lowest redshift cluster in the survey. 
Nine point sources have been detected.  
Even though this cluster has not been observed to the blank field 
confusion limit, the large number of sources might introduce extra 
uncertainties on the derived parameters. 
Three sources towards the centre of the field are just 14$''$ (just smaller 
than a  beam) from 
one another.  
One of the central sources, SMM\,J043337.8$-$131541, is coincident with the cD. 
The two other central sources, SMM\,J043337.4$-$131558 and SMM\,J043336.5$-$131547, 
are so close (just less than a beam) 
to the centre of the cD galaxy, that they are likely 
associated with cD galaxy. 
The latter of those two sources has a probable 
detection of 450\,$\mu$m emission. 
All three central sources will be excluded from the further analysis in 
this paper.  
In Figure \ref{fig:all_maps1}, source SMM\,J043338.9$-$131444 has no $S/N= $ 3 
contour as it is located in a depression in the background.  
The six sources not associated with the central cD galaxy 
are magnified by factors 1.3-1.4 and have unlensed fluxes of 3-6.3\,mJy. 
We note that two sources have more than 5 per cent of their posterior 
probability distribution below 0 mJy. 
This large number of sources in a low-z cluster field is surprising.  
Follow-up observations indicate that they are not cluster members, which 
might otherwise be expected because of the low redshift of the cluster. 
Like A478, the lens mass model is a simple model including the cluster 
potential and the cD galaxy \citep{peletier90,zabludoff90}. 
\\ \\
\underline{\it A520}\ \  The optical centre and the X-ray centre of A520 is 
not coincident, and the cluster seems to be undergoing strong dynamical 
evolution, as the cD galaxy is not located at the centre of the X-ray 
emission \citep{proust00}.  Our SCUBA map is 
about $1'$ E of the X-ray centre \citep{govoni01}. 
Four point sources have been detected 
with lensing corrected fluxes between 0.7 and 3.4 mJy. 
We note that two sources have more than 5 per cent of their posterior 
probability distribution below 0 mJy. 
The brightest source in the field, SMM\,J045409.7+025510, has a possible 
detection of 450\,$\mu$m flux. 
The mass model is based on the general cluster potential \citep{white97,carlberg96}. 
% -- map  5  (MS1054-03)
\begin{figure*}%[t!]
\begin{center}
\center{\includegraphics[width=10.7cm]{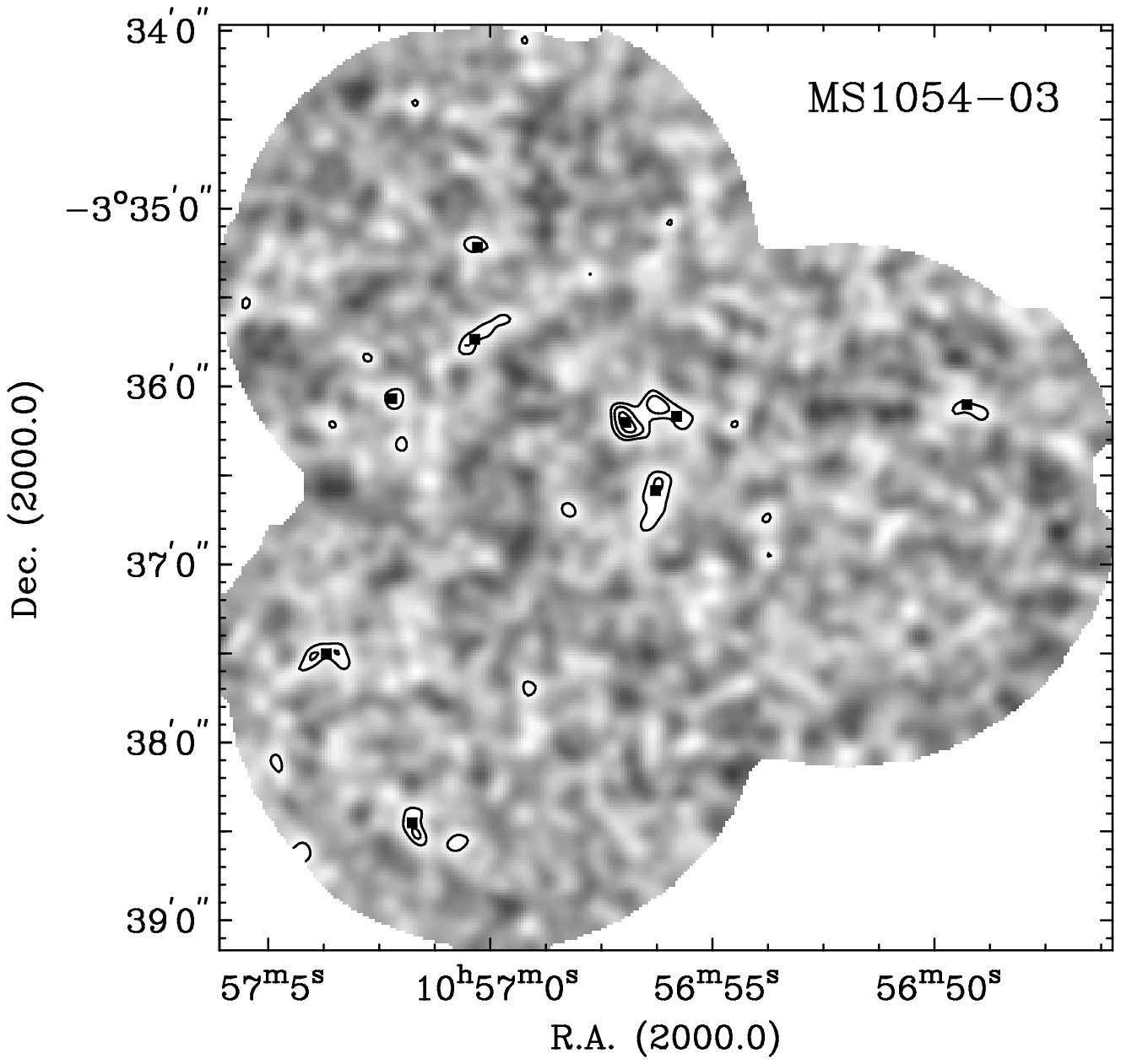}}
\caption[The maps]{
The signal-to-noise 850\,$\mu$m SCUBA map of the cluster MS1054-03.  
The overlayed contours represent $S/N= $ 3,4,5,6,7,8. 
The black boxes indicate the positions of the detected sources.  
To cover a large area of the existing multi-wavelength data (see text 
for details), we have obtained three pointings for this field.  
\label{fig:all_maps5}}
\end{center}
\end{figure*}
\\ \\
\underline{\it MS1054$-$03}\ \  For this cluster the deepest multi-wavelength 
data set exists, ranging from radio to X-ray.  It is a part of the 
FIRES-project \citep[Faint IR Extragalactic Survey;][]{franx00,forster06}, 
which includes the deepest 
near-IR imaging of a cluster taken with ISAAC at VLT.  
The area covered with ISAAC and other instruments is about 5$'\times$5$'$, 
so we decided to obtain three pointings to cover a larger area of 
the field and take advantage of the excellent data available for follow-up 
studies.  
The three pointings, which are denoted by $S$, $N$ and $NW$, according 
to the relative position cover a signification part of the FIRES field.  
The $S$ pointing is centered at the cluster centre.  
We detect nine sources with fluxes between 3.5 and 5.0 mJy.   
We note that one source has more than 5 per cent of their posterior 
probability distribution below 0 mJy, however, this source has already 
been identified with a Distant Red Galaxy at redshift $z=2.423$ 
\citep{vandokkum04,knudsen05} and thus we do not consider this a spurious 
detection.  
The largest fraction of the sources are located in the $N$ pointing, 
while the $NW$ pointing is a lot sparser.  This suggests a level 
of clustering of SCUBA sources, though the field is too small for a 
reliable clustering analysis.  
Our map is about three times deeper than the shallow map of the 
area of the $S$-pointing published by \citet{chapman02}. 
They find one source, which is off-set $\sim$\,25$''$ north of 
SMM\,J105703.7$-$033730 and not at all detected in our much deeper map. 
The gravitational lensing is not particularly strong for this cluster, 
which is partly related to the relative high redshift. 
The mass model is based on the overall cluster potential 
(\citealt{tran99}; P.\ van Dokkum, private communication). 
\\ \\
\underline{\it A1689}\ \  With $\sim$\,34 hours raw integration time 
for a single pointing and a 1$\sigma$ r.m.s.\ $\sim$\,0.7\,mJy, 
this is one of the deepest maps of the survey.  
A1689 is a cluster known to have an exceptionally large Einstein radius 
\citep[e.g.,][]{king02}.  
Because the gravitational lensing is so strong, the source plane area 
at redshift $z = $ 2.5 is only 0.3\,arcmin$^2$, i.e.\ 20 times 
smaller than the image plane area or the field of view of SCUBA.  
We detect nine SCUBA sources, and note that this field might be 
suffering from confusion due to the large number of sources.  The three 
central sources are approximately one beam from one another, and the 
same is the case for two eastern sources.   
The sources have observed fluxes between 2.6 and 5.4 mJy.  When 
correcting for the gravitational lensing magnification the fluxes are 
between 0.11 and 1.3 mJy.  
The central source SMM\,J131129.1$-$012049 has a probable detection of 
450\,$\mu$m flux.   
\\ 
Two multiply-imaged galaxies have been identified among the submm
galaxies in this field.  
SMM\,J131129.1$-$012049 and SMM\,J131134.1$-$012021 have been identified with 
the triple-imaged system 5 (as numbered in \citet{broadhurst05}), while 
SMM\,J131132.0$-$011955, SMM\,J131129.8$-$012037 and a small contribution to 
SMM\,J131134.1$-$012021 arise from either system 24 or 29, both 
quintuple-imaged galaxies.  Recent optical spectroscopy has shown that 
all these galaxies have redshifts $\sim 2.5$ (Knudsen et al., in prep., 
and Richard et al., in prep.).  A detailed analysis of the identification 
will be discussed in a future paper along with additional multi-wavelength 
follow-up.  
The mass model will be presented in detail in Richard et al.\ ({\it in prep.}) 
and Limousin et al.\ ({\it submitted}). 
%
% -- map-mosaic 2
\begin{figure*}%[t!]
\begin{center}
\center{\includegraphics[width=12.9cm]{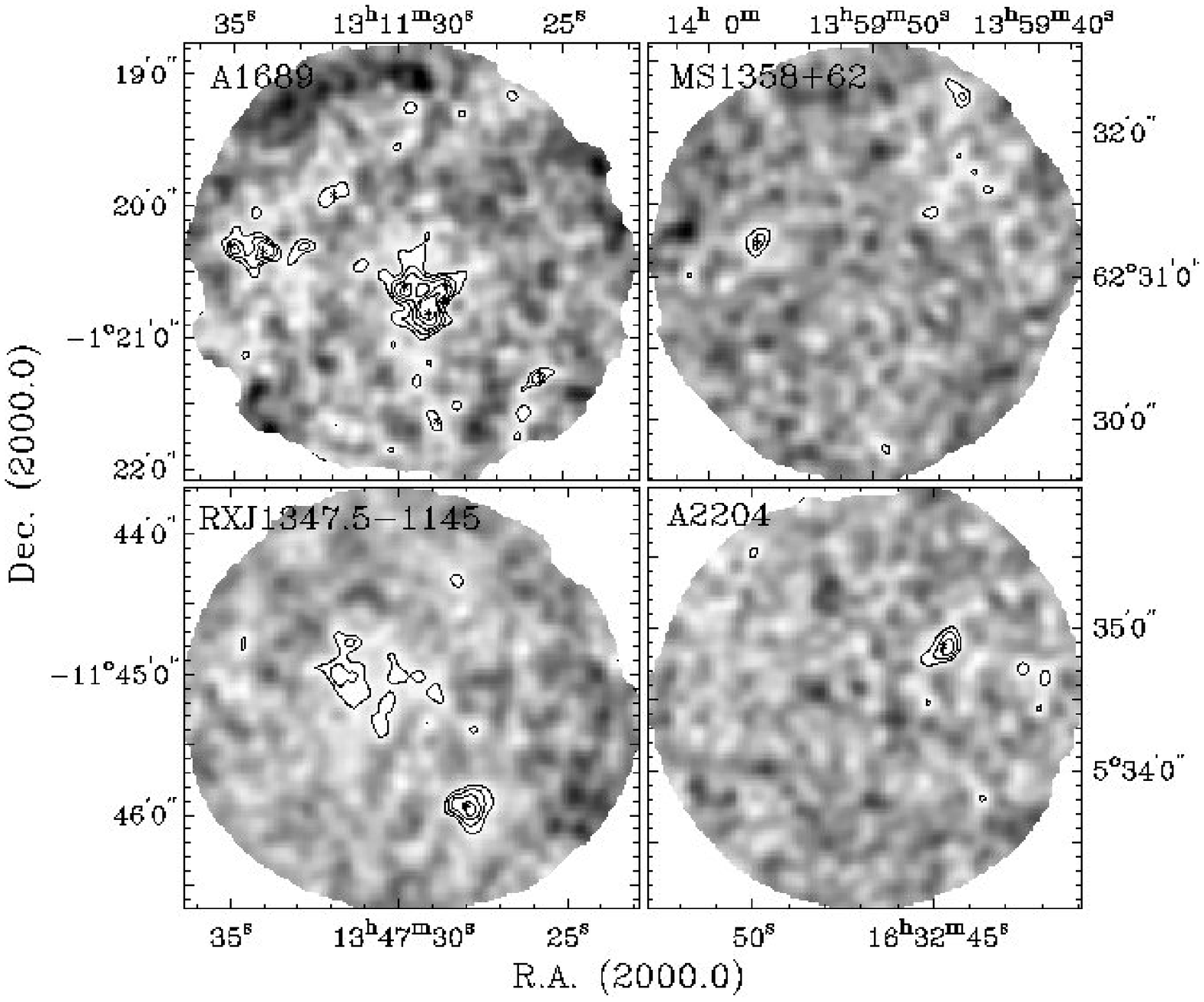}}
\caption[The maps]{
The signal-to-noise ($S/N$) 850\,$\mu$m SCUBA maps of the clusters 
A1689, MS1358+62, RX\,J1347.5-1145 and A2204.  The overlayed contours 
represent $S/N = $ 3,4,5,6,7,8.  The stars $\star$ indicate the positions 
of the detected sources.  
\label{fig:all_maps2}}
\end{center}
\end{figure*}
\\ \\
\underline{\it RX\,J1347.5$-$1145}\ \  This is the most X-ray-luminous cluster 
known \citep{allen02}. 
In this field we detect one source, SMM\,J134728.0$-$114556, which has 
$S_{850} = $ 16.2$\pm$3.1\,mJy and $S_{450} = $ 98.7$\pm$29.6\,mJy.  
The source is strongly lensed and has an unlensed flux of 4.5 mJy. 
Furthermore, a large, extended source near the cluster centre is present. 
This is the Sunyaev--Zel'dovich effect reported in \citet{komatsu99}. 
The mass model is based on \citet{cohen02}. 
\\ \\
\underline{\it MS1358+62}\ This 
map is relatively empty with only one detection 
SMM\,J135957.1+623114, with $S_{850} = $ 6.8$\pm$1.3\,mJy; 4.4 mJy 
after correcting for the lensing.  
These data were first obtained to study the strongly lensed, redshift 
$z= $ 4.92 galaxy MS1358+62-G1 \citep{franx97}, 
which however was not detected \citep{vanderwerf01}.  
We here find a 3$\sigma$ upper limit for G1 of $S_{850} < $ 4.8\,mJy. 
A detailed mass model describes the potential for this cluster 
\citep{franx97,santos04}. 
\\ \\
\underline{\it A2204}\ \  This field has the shallowest SCUBA 
observations of the whole survey. 
One point source, SMM\,J163244.7+053452, has been detected at a $S/N = $ 4.9 
with an observed flux of $S_{850} = $ 22.2$\pm$5.7\,mJy 
making it the second brightest source in the catalogue. 
This source is lensed by more than a factor 3, resulting in a corrected 
flux of $\sim$\,7\,mJy.  
\\ \\
\underline{\it A2218}\ \  Together with A1689 and the NTT Deep Field, 
this field is the deepest data taken for the survey.  
The data for this field cover an area corresponding to more 
than two pointings.   
The data was used as a case study for the source 
extraction method, the Mexican Hat Wavelets algorithm, applied for this
survey \citep{knudsen06}. 
% -- map  4  (A2218)
\begin{figure}%[t!]
\begin{center}
\center{\includegraphics[width=8.6cm]{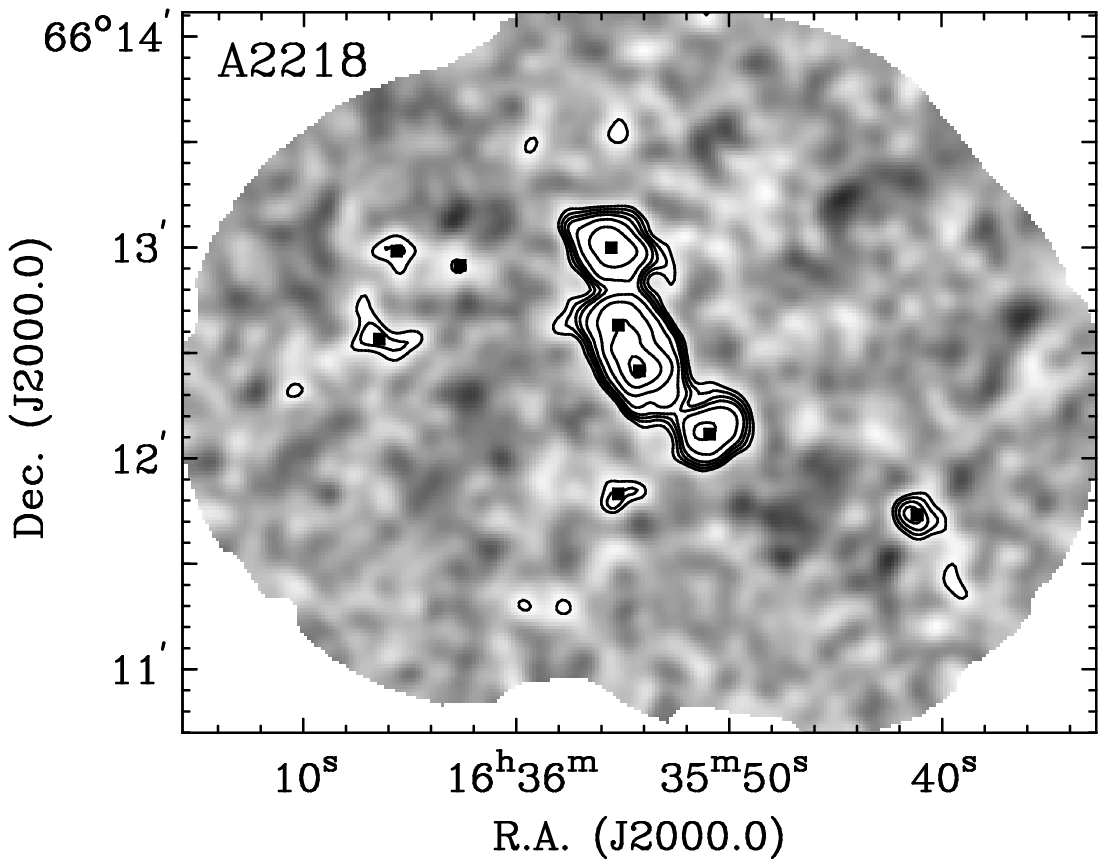}}
\caption[The maps]{
The signal-to-noise 850\,$\mu$m SCUBA map of the cluster A2218.  
The overlayed contours represent $S/N= $ 3,4,5,6,8,12,18,24. 
The black boxes indicate the positions of the detected sources.  
\label{fig:all_maps4}}
\end{center}
\end{figure}

In this field nine sources were detected.  
The three source, SMM\,J163550.9+661207, SMM\,J163554.2+661225 and SMM\,J163555.2+661238, 
have been identified as the same, multiply-image source at redshift $z= $ 2.516 
\citep{Kneib_a2218}, and is referred to as SMM\,J16359+6612. 
The source SMM\,J163555.2+661150, which is detected both at 850\,$\mu$m 
and 450\,$\mu$m, is coincident with a known galaxy, \#289, with 
redshift $z = $ 1.034 \citep{pello92} and is also detected 
at 15\,$\mu$m with ISOCAM \citep[e.g.,][]{metcalfe03}. 
The relatively bright source SMM\,J163541.2+661144 is detected at both 
850\,$\mu$m and 450\,$\mu$m.   
The sources in A2218 have observed fluxes between 2.8 and 16.1 mJy. 
The lensing corrected fluxes are 0.4--6.1\,mJy. 
The mass model is based on \citet{kneib96,ellis01,Kneib_a2218}. 
% -- map-mosaic 3
\begin{figure*}%[t!]
\begin{center}
\center{\includegraphics[width=12.9cm]{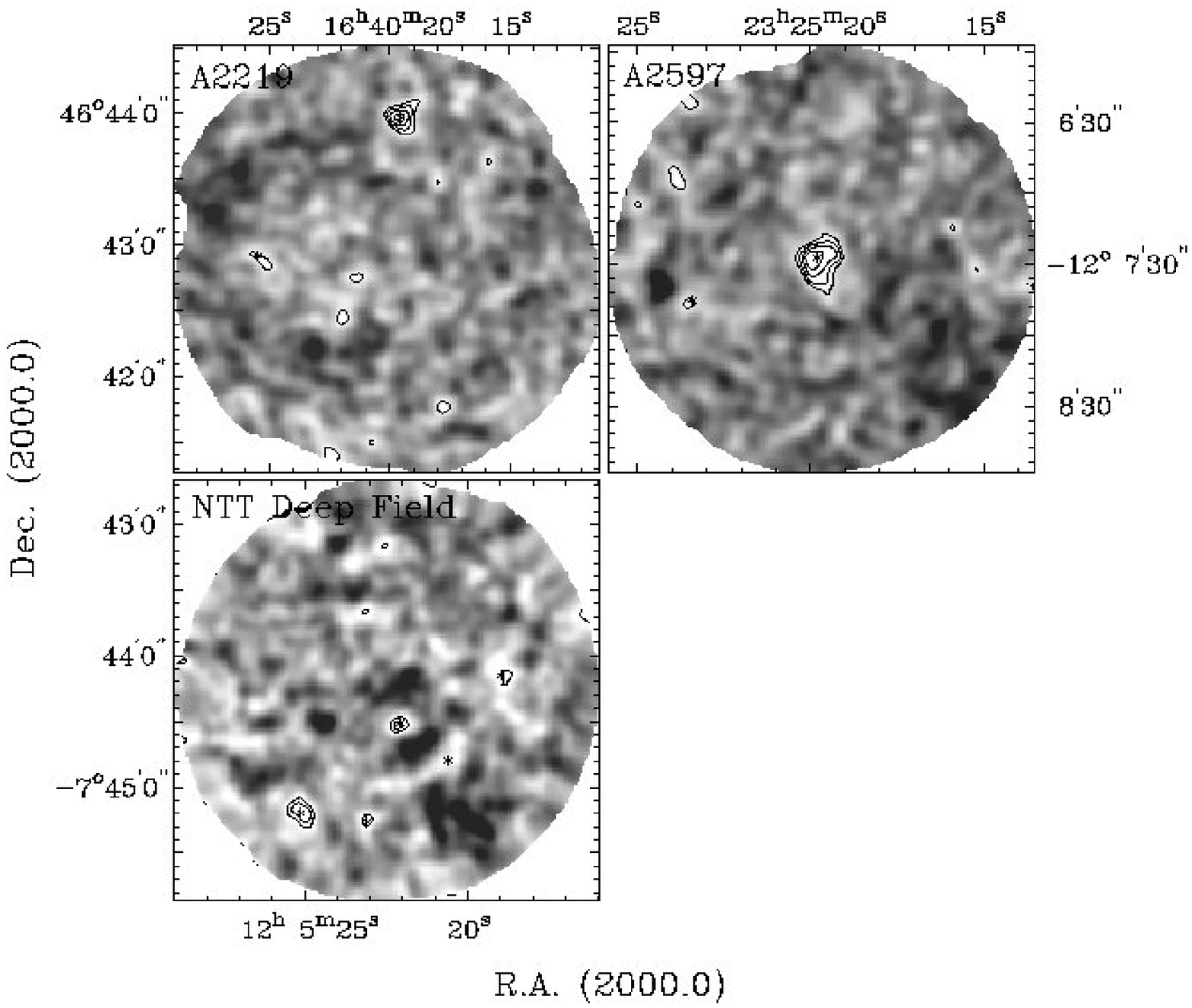}}
\caption[The maps]{
The signal-to-noise ($S/N$) 850\,$\mu$m SCUBA maps of the clusters 
A2219 and A2597 and of the blank field NTT Deep Field.  
The overlayed contours represent $S/N= $ 3,4,5,6. The stars $\star$ indicate 
the positions of the detected sources. 
\label{fig:all_maps3}} 
\end{center}
\end{figure*}
\\ \\
\underline{\it A2219}\ \  In this field we detect two point sources that 
both have possible detections of 450\,$\mu$m flux.  
This field was also a part of the \citet{chapman02} sample, 
though their observations are shallower.  The source SMM\,164019.5+464358
agrees well with their finding.  In \citet{chapman02} 
the source C\,SMM\,J16404+4643 has an upper limit at 850\,$\mu$m, 
while being detected at 450\,$\mu$m.  We do not get a significant 
detection of the source, but MHW does suggest a $S/N\sim $ 1.9 detection, 
which corresponds to a flux of $\sim$\,2.3\,mJy; the 3$\sigma$ upper 
limit is $<$\,3.6\,mJy. 
We note that one source has more than 5 per cent of its posterior 
probability distribution below 0 mJy. 
For the source B\,SMM\,J16403+46437 MHW suggests a $S/N\sim $ 2.3 detection. 
The source D\,SMM\,J16404+4644 is within the edge region which have 
trimmed from the map.   
In our map 
a positive fluctuation is present, though it does not have the characteristics  
of a significant 850\,$\mu$m detection.   
Furthermore, MHW also finds a $S/N\sim $ 2.9 source at $\alpha,\delta$ = 
16$^{\rm h}$40$^{\rm m}$22$^{\rm s}$,+46$^\circ$42$'$25$''$. 
The mass model is described in \citet{smith05}. 
\\ \\
\underline{\it A2597}\ \  In this field, two point sources were detected with 
$S/N > $ 3. 
The brightest source, SMM\,J232519.8$-$120727, is a 12 mJy source located in 
the centre of the map and is coincident with the cD galaxy of the cluster, 
which is a well-known AGN \citep[e.g.,][]{mcnamara99}.  
The cD galaxy is excluded from the rest of the analysis in this paper. 
The other source, SMM\,232523.4$-$120745, is a 5 mJy source, which 
also has detected 450\,$\mu$m flux. 
We note that this source has more than 5 per cent of their posterior 
probability distribution below 0 mJy. 
Hence, if this is indeed a spurious source, then no high-$z$ background
sources were detected in this field.  
The mass model includes both the overall potential of the cluster and 
that of the cD galaxy \citep{smith90,wu00}. 
\\ \\
\underline{\it NTT Deep Field}\ \  This field, the blank field of the survey 
\citep{arnouts}.  is one of the deepest fields of the survey.  
Five sources have been detected with fluxes between 3 and 4 mJy.  
None of them have detected 450\,$\mu$m flux. 
In Figure \ref{fig:all_maps3}, source SMM\,J120520.6$-$074448 has no $S/N= $ 3 
contour as it is located in a depression in the background.  
A 1.2 millimeter map of the NTT Deep Field has been obtained with 
the Max-Planck Millimetre Bolometer (MAMBO) covering a larger area of 
the NTT Deep Field than the SCUBA map presented here 
\citep{dannerbauer02,dannerbauer04}.  Considering that 1.2\,mm 
probes a part of the modified blackbody where the flux is fainter compared 
to the 850\,$\mu$m, the MAMBO map is a bit shallower than the deep 
SCUBA map.  Two MAMBO sources are covered by the SCUBA map.  The source 
MM\,J120517$-$0743.1 is very close to the edge of the SCUBA map, where there 
are no indications of a source.  The source MM\,J120522$-$0745.1, which 
has a radio detection, is only 6$''$ from the submm source SMM\,J120523.1$-$074516. 
The radio detection is coincident with the submm source.

\label{lastpage}

\end{document}